\documentclass[useAMS,usenatbib]{mn2e}
\usepackage{graphicx}
\usepackage{hyperref}
\usepackage{amsmath}
\usepackage{amssymb}
\usepackage{mathtools}
\usepackage{bm}
\usepackage{cleveref}

\title[Axially symmetric equations for pulsar rotation]
{Axially symmetric equations for differential pulsar rotation with superfluid entrainment }
\author[M.Antonelli \& P.Pizzochero]{
M.~Antonelli$^{1,2}$\thanks{E-mail: marco.antonelli@unimi.it},
P.~M.~Pizzochero$^{1,2}$\\  
$^{1}$Dipartimento di Fisica, Universit\`a degli Studi di Milano, Via Celoria 16, 20133 Milano, Italy\\ 
$^{2}$Istituto Nazionale di Fisica Nucleare, sezione di Milano, Via Celoria 16, 20133 Milano, Italy}
\begin{document}
\date{Submitted to MNRAS in Feb-2016.}
\pagerange{\pageref{firstpage}--\pageref{lastpage}} \pubyear{2016}
\maketitle
\label{firstpage}
\begin{abstract}
In this article we present an analytical two-component model for pulsar rotational dynamics. 
Under the assumption of axial symmetry, implemented by a paraxial array of straight vortices that thread the entire neutron superfluid, 
we are able to project exactly the 3D hydrodynamical problem to a 1D cylindrical one.
In the presence of density-dependent entrainment the superfluid rotation is non-columnar: 
we circumvent this by using an auxiliary dynamical variable directly related to the areal density of vortices.
The main result is a system of differential equations  that take consistently  into account
the stratified spherical structure of the star, the dynamical effects of non-uniform entrainment, 
the differential rotation of the superfluid component and its coupling to the normal crust. 
These equations represent a mathematical framework in which to test quantitatively the macroscopic 
consequences of the presence of a stable vortex array, a working hypothesis widely used in glitch models. 
Even without solving the equations explicitly, we are able to draw some general quantitative conclusions; 
in particular, we show that the reservoir of angular momentum (corresponding to recent values of the pinning forces), 
is enough to reproduce the largest glitch observed in the Vela pulsar, provided its mass is not too large.
\end{abstract}
\section{Introduction}
\label{sec:into}
The detection of glitch events in many isolated radio pulsars indicates that a large amount of angular momentum must 
be first stored and then exchanged between an internal component and the rigid observable crust.
Thus a neutron star has to be comprised of (at least) two components, spinning at slightly different velocities,
that interact via some type of mutual friction in order to permit such an accumulation and exchange of angular momentum. 
Observations of very long post glitch relaxations indicate that one of these components should be superfluid, whereas
the spin-up is so fast that it is still unresolved (less than one minute).
 
With this in mind, these recurring period instabilities are usually explained in terms of superfluid vortex dynamics \citep{AI75},  
a scenario where pinning of vortex lines to the crustal lattice freezes part of the angular momentum  and 
glitches are the consequence of a vortex avalanche, a catastrophic event in which many vortices are suddenly expelled 
from the superfluid bulk and transfer their angular momentum to the crust.
%thanks to the combined effect of the drag and Magnus force. 
The nature of the event that triggers an avalanche is still debated:
it could be a starquake \citep{R76,RZ98}, a hydrodynamic instability \citep{GA09} or 
a manifestation of a self-organized critical system \citep{MP08}. 
Despite the variety of existing studies, however, glitch theory lacks consistent macroscopic models
whose quantitative outputs can be compared with observational data: for example the need to take into account the spherical 
and non-uniform structure of the star in order to evaluate the angular momentum reservoir was pointed out only recently in 
the framework of the static ``snowplow model'' \citep{P11}.
Since the pioneering two-component model of \cite{BP69}, detailed dynamical models have evolved  in order 
to incorporate in the dynamical equations different subtle effects of multi-fluid hydrodynamics, such as superfluid entrainment \citep{C89, CL94, C08}.
Although the local hydrodynamical equations are well known, 
the great majority of studies of glitches have been carried out in the quite unphysical approximation of  a cylindrical, 
uniform-density and rigidly rotating star: as pointed out by \cite{AG12}, 
there has been little progress on making the pulsar glitch models quantitative.

In this work we address the problem of how to construct a simple but fully consistent model for pulsar glitches. 
By {\em consistent} we mean that, starting from well stated hypothesis, we provide a coherent prescription of how to implement 
all the available microphysical information in our macroscopic set of equations. 
This permits to account for the fact that  radially the star has a non-uniform structure which in turn implies that the rotation 
of the superfluid must be {\em differential}, even in the simplest situation of parallel and straight vortices.

Our study is in the framework of the two-component model: a charged, rigid  p-component is spun down electromagnetically,
while a non-observable n-component comprised of superfluid neutrons reacts to the spin-down via some type of mutual friction. 
The hydrodynamical formalization of this multi-fluid problem is nowadays quite clear and developed 
but leads to a computationally difficult 3D problem [see for example \cite{AC07} and \cite{HM15} for recent reviews 
and \cite{PM06} for hydrodynamical simulations in Couette flow geometry]. 
We can reduce it to a simpler 1D model by making two widely used assumptions: axial symmetry and straight vortices.
On the one hand, the assumption of axial symmetry around the rotational axis of the pulsar is a widespread working hypothesis
and allows to reduce the dimensionality of the problem. In this way, however, we loose the possibility to study the nature of the trigger event,
since probably it is related to some non-trivial effect of the fully three-dimensional problem. 
On the other hand, the assumption of a paraxial array of straight vortices which resist bending is justified by the collective 
rigidity of vortex bundles in coherent motion: it is consistent with and required by the assumption of  axisymmetry in a rotating superfluid, 
as discussed in a seminal paper by \cite{RS74}. 
This allows explicit integration along the vortex lines, projecting the spherical star onto its equatorial plane. 
In this way, however, we rule out a priori turbulent motion and its consequences on the mutual friction between the two 
components \citep{AS07}: this is a limit of the model to be kept in mind.

In this paper we develop a dynamical model for superfluid differential rotation with entrainment: 
this is a new scenario since in the literature we can only find rigid models with uniform entrainment [as for example \cite{SP10}] 
or differential models without entrainment [as done in \cite{HP12}].
The recent large values found for entrainment in the crust indicate that such a phenomenon 
cannot be ignored in the physics of pulsar glitches \citep{C12,C13,AG12}. 
Moreover, entrainment turns out to be strongly density-dependent, as shown in section \ref{sec:num-integration}.
Thus an important part of our work is devoted to dealing with the complication that in the presence of density-dependent entrainment the hydrodynamics 
cannot be {\em columnar}, even in the extremely simple case of straight and parallel vortices: because of non-uniform entrainment, 
the angular velocity of the n-component will depend on the coordinate along the axis of rotation. 

In the next section we will describe our working hypothesis and how to handle the entrainment effect when there are
macroscopically long vortices that pass trough spherical layers of different density and composition. 
In the absence of a layer of normal matter between the core and the inner crust, the normal vortex cores should pass continuously through the phase transition between the S-wave superfluid in the crust and the P-wave superfluid in the core \citep{ZS04}; therefore, we take vortices which thread the entire neutron superfluid. 
We stress, however, that this is only a possibility and that it is straightforward to generalize our model to account for two different superfluid components (as long as we keep the assumption of straight vortices).
 The main result of the paper are the equations of motion of the system presented at the beginning of section \ref{sec:macro-eqs}; they
are written in a form that does not depend explicitly on the presence of entrainment or by the fact that 
vortices pass trough the crust-core interface.
These dynamical equations generalize and extend the previous work of 
\cite{HP12} to the case of density-dependent superfluid entrainment and (with a suitable choice of the mutual friction term) provide a dynamical realization of the static "snowplow model" introduced by \cite{P11}.
Thus the present work sets a rigorous mathematical framework in which to test quantitatively the macroscopic consequences of the presence of a stable array of parallel vortices, a  working assumption widely used in almost all existing glitch models.
In the last two sections, we give a numerical example of our results, using consistently some of the microphysical 
information that can be found in the literature and employing two different equations of state to describe dense matter.
This also permits us to give estimates of the superfluid angular momentum reservoir and of the spin-up timescales as a function of the stellar mass. 
A concluding section summarizes the physical relevance of the paper for future numerical studies of pulsar glitches.

\section{Description of the hydrodynamical model}
\label{sec:construction-hydro-model}
In this section we derive the dynamical equations needed to describe the differential rotation 
of a pulsar in the framework of the two component model with superfluid entrainment.
As already discussed in the introduction, in the presence of density-dependent entrainment 
the superfluid velocity field cannot be columnar, even if boundary effects and turbulence are totally neglected.
To overcome this, we will not consider directly the velocity of the superfluid as a dynamical variable, 
but rather an auxiliary variable directly associated to the vortex line density.  
\subsection{Model assumptions}  
\label{subsc:hyp} 
\begin{figure}
    \centering
    \includegraphics[width=.45 \textwidth]{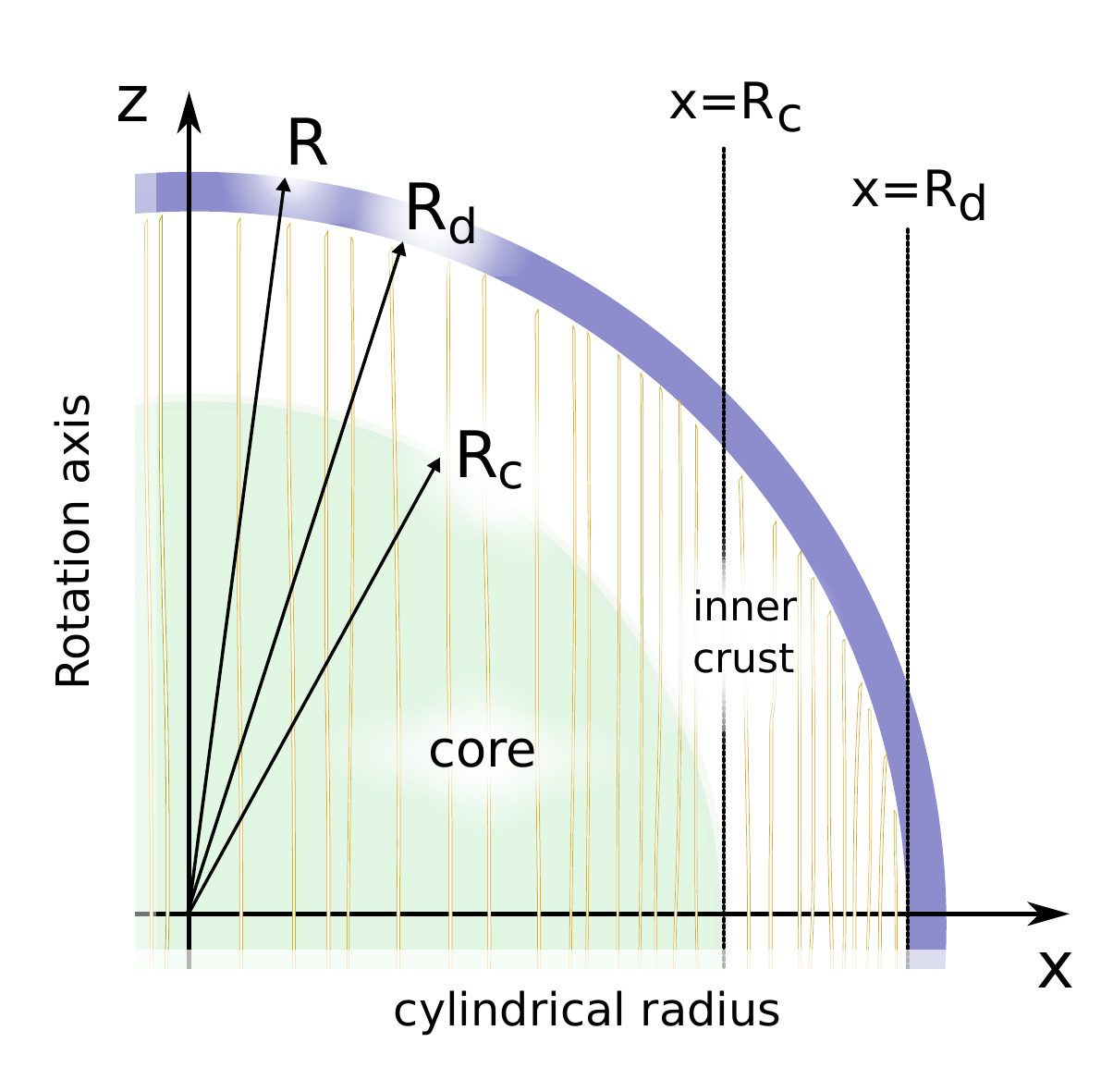}
    \caption{Sketch of the stellar structure (out of scale), with the geometrical definitions used in the model. In the cylindrical shell
        $R_c<x<R_d$ the vortex lines are completely immersed in the inner crust. The outer crust (the thin blue layer in the figure)
        is part of the p-component.
    }
    \label{fig:structure}
\end{figure}

In the following we describe the set of assumptions, some of which can be generalized at a later time,
that are used to construct our model:   
\begin{enumerate}
\item The star is spherical, spinning around the z-axis (namely there is no precession), with inner-to-outer crust interface at $R_d$ (the radius at which the neutron drip starts).
In order to simplify the fully 3D hydrodynamical problem, we will use the hypothesis of axial symmetry around 
the rotation axis of the star. To implement this symmetry, the slow-down of the pulsar is given by an external braking torque with 
constant direction $\bm{T}_{ext} = -{T}_{ext}\hat{\bm{e}}_z$ that acts on the star. 
\item We consider only one superfluid component (the n-component) inside the spherical volume of radius $R_d$.
This superfluid component is threaded by straight vortices parallel to $\hat{\bm{e}}_z$ (as in Fig \ref{fig:structure}),
each carrying a quantum of circulation $\kappa=h/(2m_n)$ (with $m_n$ the nucleon mass). 
We discuss further this point at the end of this section. 
\item In the absence of a layer of normal matter between the core and the inner crust \citep{ZS04}
there is no clear distinction between the crustal superfluid and the superfluid in the core
and we assume that normal matter vortex cores  pass 
continuously through the phase transition between the S-superfluid to the P-superfluid.
This is the simplest possibility but it can be eventually modified
by considering superfluidity only inside a spherical shell of radii $R_c$ and $R_d$. Although the scenario of continuous vortex lines was already suggested by \citet{R76} as an alternative to the scenario of distinct vorticity in the crust and the core, in the subsequent literature only the latter assumption has been implemented until the "snowplow model" of \citet{P11}.
\item The rigid crust and the charged component (electron and proton fluids) are locked into the strong magnetic field of
the neutron star on very short timescales (the p-component). 
This is justified as long as protons (or ions in the crust) and electrons are locked on timescales much shorter than the dynamical timescales we are interested in \citep{E79}.
Using cylindrical coordinates $(x,\theta, z)$, the velocity field of this rigid p-component 
is $\bm{v}_p = x \Omega_p \hat{\bm{e}}_\theta$, which is related to the observed pulsar period $P=2\pi \Omega_p^{-1}$. 
\item As usual in the two-fluid approach, the presence of vortices induces a mutual friction between the n-component and the p-component, 
due to the presence of internal dissipative channels that can be modeled as drag forces acting on vortex lines \citep{AS06}. For each
dissipative process (``d.p.'') occurring at a given density $\rho$, we need a corresponding drag parameter 
$\eta_{\text{d.p.}}(\rho)$ that gives the intensity of the associated drag force on vortices.
\item Vortex lines can also pin to the different inhomogeneities present in the star, thus we need a quantity that (for different 
densities and pinning mechanisms) tells us to which extent the vortices can be blocked. We assume that pinning can be modeled 
by introducing a ``pinning force per unit length (of vortex line)" $f_p(\rho)$, which expresses the strength of the vortex-inhomogeneity 
interaction, suitably mediated over a mesoscopic region at density $\rho$. 
%Since vortex lines pass trough the crustal lattice we will also consider the possibility of pinning. This will 
%be modeled by introducing a mesoscopic pinning force (per unit length) $f_p(\rho)$ that can be interpreted as the local strength of
%the vortex-lattice interaction. 
\end{enumerate}
Given this set of assumptions, we are far from describing the real hydrodynamical problem 
of a spinning neutron star interior; nonetheless this is 
a first attempt to construct a differential model for the rotational dynamics that encodes the non-uniform 
structure of the star and the entrainment effect in a consistent way, being at the same time computationally easy to implement. 
In particular, the hypothesis of an array of straight vortex lines and its validity to describe neutron star rotation deserve some further discussion.
On the one hand,  the tension of a single line is much smaller than 
the hydrodynamical force acting on a vortex, which suggests bending and twisting of lines so that vortices would actually form a tangle: turbulence is thus expected to develop in neutron stars interiors and indeed, following a criterion for the onset of turbulence originally devised for liquid Helium, \citet{AS07} find that the core is susceptible to become turbulent. On the other hand, as also explicitly remarked by the authors, this study does not account for the stabilizing effect of rotation in suppressing turbulence: the increased effective tension of a bundle of 
vortices as compared to a single vortex line resists bending at the macroscopic  scale, so that the array of vortex lines can remain parallel to the rotation axis and thus implement a superfluid flow which is consistent with a generalisation of the Taylor-Proudman theorem to axisymmetric spinnig-down frictionless fluids \citep{RS74}. 
Since superfluid turbulence is a well known phenomenon in laboratory experiments, it is likely 
to  develop in some regime also in neutron stars, but its effect on macroscopic hydrodynamics and its relevance for glitch models are still debated: further effort along these lines is required to take the modelling of glitches to the next level. As it stands, studies based on the widely used approximation of straight vortices, that is also likely to be valid in some  dynamical regimes, are still meaningful and physically relevant.
\subsection{Some preliminary definitions}
The microphysical input that describe the stellar structure are functions of the density: 
to study the rotational dynamics of the two components we need 
the superfluid neutron fraction $x_n(\rho)$ (that is zero below the drip density),
the normal fraction $x_p(\rho)=1-x_n(\rho)$, the entrainment parameter $\epsilon_n(\rho)$,
the pinning force per unit length $f_{p}(\rho)$ and the drag parameter $\eta(\rho)$ arising from 
the dissipative processes acting on vortex lines
\begin{equation}
\label{eq:gen-drag}
\eta(\rho)= \sum_{\text{``d.p.''}} \eta_{\text{d.p.}}(\rho)
\end{equation}
where, for example, a dissipative process can be the  ``electron scattering off vortex cores'' \citep{AL84,BE89}.
Given an equation of state (EOS) for the stellar matter and solving the Tolman-Oppenheimer-Volkov (TOV) equations,
all these profiles are easily parametrized as functions of the spherical radius $r$.

It is useful to introduce the following integrations along a straight line 
placed at distance $x$ from the rotation axis:
\begin{align}
\label{eq:a}
a(x)& = 2 \int_{0}^{z(x)} \! dz \,\, \rho_n(r)
\\
\label{eq:b}
b(x)& = 2 \int_{0}^{z(x)} \! dz \,\,  \frac{\rho_n(r)}{1-\epsilon_n(r)}
\\
\label{eq:d}
d(x)& = \frac{2}{\kappa} \int_{0}^{z(x)} \! dz\,\,  \eta(r) \, ,
\end{align}
where $\rho_{n,p}(r) = x_{n,p}(r) \rho(r)$ and $z(x)=\sqrt{R^2-x^2}$.
In the above integrations the spherical radius $r$ is used as a shorthand for $\sqrt{x^2+z^2}$.
We will also use a ``cylindrical mean'', defined as
\begin{equation}
\label{eq:radial-mean}
\langle f(...) \rangle = \int_0^R \! dx \, g(x)\,f(x,...) \, ,
\end{equation}
with normalized weight 
\begin{equation}
\label{eq:g-def}
g(x)=2 \pi\,x^3\,b(x)/I_v \, .
\end{equation}
The normalization factor   
\begin{equation}
\label{eq:iv}
I_v = \frac{8 \pi}{3} \int_0^R \! dr \, r^4 \, \frac{\rho_n(r)}{1-\epsilon_n(r)} \, 
\end{equation}
reduces to the moment of inertia of the n-component if there is no entrainment ($\epsilon_n=0$).
Since in all the above definitions the integrand functions are proportional to $x_n$, the stellar radius $R$ can be 
replaced by $R_d$.
\subsection{Conservation of vorticity}
\label{subsc:conservation-vorticity}
To derive the dynamical equations we could start from the well known equations
for multifluid hydrodynamics, as in \cite{HP12}, but here we prefer to present a simple 
constructive derivation. The first part of the following argument is not totally new \citep{AA84}, but the effect
of  entrainment and of the radial stellar structure have not been considered previously.
Following the variational formalism originally developed by \cite{C89}, the momentum per particle is
a linear superposition of the local kinematic velocities of the two components:
\[
\bm{p}_n(\bm{x}) \,=\,
m_n \left[\,\left(1-\epsilon_n(\bm{x})\right) \bm{v}_n(\bm{x})\,
+\,\epsilon_n(\bm{x})\bm{v}_p(\bm{x})\,\right] \, ,
\]
where we adopted the notation for entrainment discussed by \cite{PC02}.
Now the Feynman relation takes the form of Bohr-Sommerfeld quantization rule:
\begin{equation}
\label{eq:feynman-rel}
\int_\gamma \bm{p}_n\cdot d\bm{l} = \frac{h \, N(\gamma)}{2}
\end{equation}
where $N$ is the number of vortex lines enclosed by the contour of integration
$\gamma$ and the factor of two accounts for Cooper pairs.
This expression can be made local by introducing an areal density of vortices $n_v$ 
at any point $(x,\theta)$ on the equatorial plane. This quantity is well defined because of  
assumption (ii) and depends only on the cylindrical radius $x$ thanks to assumption (i).
Thus, by taking the circulation around a ring of radius $x$, the Eq. \eqref{eq:feynman-rel} reads
\begin{equation}
\label{eq:fey-nv}
\int_x \bm{p}_n\cdot d\bm{l} =  h \, \pi \! \int_0^x \! dy \, y \, n_v(y) \, .
\end{equation}  
Translation of the integration contour along the $z$-axis 
always encircle the same number of vortex lines and thus the above relation implies
that the azimuthal component of the momentum (denoted by $p_n$) is a function of the cylindrical radius only, namely
% $\bm{p}_n(\bm{x}) = p_n(x) \bm{e}_\theta + p_x(\bm{x}) \bm{e}_x + p_z(\bm{x}) \bm{e}_z $.
$\bm{p}_n(\bm{x}) \cdot \hat{\bm{e}}_\theta = p_n(x)$.
On the one hand, it is convenient to introduce an auxiliary, columnar angular velocity defined by
\begin{equation*}
\Omega_v(x) = \frac{p_n(x)}{m_n\,x} \,\, .
\end{equation*}
From Eq. \eqref{eq:feynman-rel} this quantity obeys the usual Feynman relation:
\begin{equation}
\label{eq:omegav}
\Omega_v(x)=\frac{\kappa\, N(x)}{2\pi\,x^2} \, \, .
\end{equation}
On the other hand, the  angular velocity of the n-component is non-columnar due to the density dependence of entrainment and is given by
\begin{equation}
\label{eq:omegan}
\Omega_n(x,z) = \frac{\kappa}{1-\epsilon_n(r)}\frac{N(x)}{2\pi\,x^2}-\frac{\epsilon_n(r)}{1-\epsilon_n(r)} \Omega_p \, .
\end{equation}
Stokes' theorem applied to Eq. \eqref{eq:fey-nv} implies that the density of vortex lines $n_v$ is 
proportional to the component of the vorticity directed along the normal to the equatorial plane
%NOTA: VEDI TEOREMA DI STOKES SU KLEINERT, MULTIVALUED FIELDS, ORIGINALE! 
\begin{equation}
\label{eq:fey-loc}
n_v(x) = \frac{ (\nabla \times \bm{p}_n) \cdot \hat{\bm{e}}_z }{\kappa\, m_n} = \frac{1}{\kappa} (2\Omega_v(x) + x \partial_x \Omega_v(x)) \, \, .
\end{equation}
Exchange of angular momentum between the superfluid and the p-component is possible only if we allow for radial motion of vortices.
Let $\bm{v}_L = (v_L^x , v_L^\theta, 0 )$ the local mean velocity of vortex lines in the ``laboratory'' rest frame. 
The conservation of the number of intersections of the vortex lines with the equatorial plane implies the following 
continuity equation
\begin{equation*}
\partial_t n_v + \frac{1}{x} \partial_x ( x \,n_v \,v_L^x) = 0 \, \, . 
\end{equation*}
Taking the temporal derivative of Eq. \eqref{eq:feynman-rel} and
using the above expressions (in particular the continuity equation), it's easy to obtain 
\begin{equation}
\label{eq:first-eq-gen}
\partial_t \Omega_v(x,t) = -\left( 2\Omega_v(x,t) + x\partial_x \Omega_v(x,t) \right)\frac{v_L^x(x,t)}{x} \,\, .
\end{equation}
This is the first dynamical equation of the model: once the radial velocity $v_L^x(x,t)$ is given, it expresses the conservation of the vorticity
of the superfluid component,  in terms of a first order partial differential equation for the auxiliary variable $ \Omega_v(x,t)$.
\subsection{Total angular momentum balance}
\label{subsc:total-ang-mom-bal}
%Our first hypothesis tells us that $\dot{L}^z_n+\dot{L}^z_p=-T_{ext}$:
The isolated pulsar slows down under the action of the external braking torque due to radiation emission, namely
\begin{equation}
\label{eq:second-eq-general}
\int \! d^3 \! x \,\,\bm{x}\times \left[ \rho_n(r)\dot{\bm{v}}_n(x,z) +  \rho_p(r)\dot{\bm{v}}_p(x)   \right] \, = \,\bm{T}_{ext} \, .
\end{equation}
With the substitution
\[
\bm{v}_n(x,z) = x\frac{\Omega_v(x) - \epsilon_n(r) \Omega_p }{1-\epsilon_n(r)}\, \hat{\bm{e}}_\theta 
\]
and using the well known property $\rho_n(r)\epsilon_n(r)=\rho_p(r)\epsilon_p(r)$, 
after a simple calculation Eq. \eqref{eq:second-eq-general} takes the form
\begin{equation}
\label{eq:second-equation}
\partial_t \Omega_p(t) =-\frac{T_{ext}}{I_p} - 
\frac{2\pi}{I_p}\int_0^R \! dx\,x^3\,\partial_t \Omega_v(x,t)\,b(x) \,\, .
\end{equation}
Here we used Eq. \eqref{eq:b} and $I_p$ is defined as
\begin{equation}
\label{eq:ip}
I_p  = \frac{8 \pi}{3} \int_0^R \! dr\, r^4\,\rho_p(r)\frac{1-\epsilon_n(r)-\epsilon_p(r)}{1-\epsilon_n(r)} 
\end{equation}
and in the absence of entrainment it reduces to the moment of inertia of the normal component; from equation \eqref{eq:second-equation}
it is apparent that $I_p$ is the moment of inertia associated to the p-component in the presence of entrainment.
This quantity is well defined since the integrand is always positive, thanks to the physical constraints on the entrainment parameters
\[
\epsilon_n < x_p  \,  , \quad \epsilon_p < 1-x_p = x_n \, ,
\]
valid in any layer of the star \citep{CH06}. 
Obviously  entrainment  cannot change the total moment of inertia of the star 
\[
I \, = \, \frac{8 \pi}{3}\int_0^R \! r^4 \rho(r) \,dr 
\]
and, since the relation
\[
I_p + I_v \, = \, I      
\]
holds, we conclude that $I_v$ defined in Eq. \eqref{eq:iv} is indeed the correct moment of inertia
associated to the auxiliary variable $\Omega_v$.
%\par 
%We stress that $\Omega_p$ is the pulsation that we detect when observing a pulsar, whereas 
%$\Omega_v$ is a convenient dynamical variable related with the unobservable configuration of vortex lines via Eq. \ref{eq:omegav}.
%
\subsection{Relation between vortex velocity and lag}
\label{subsc:relation-vortex-vel-lag}
Equations \eqref{eq:first-eq-gen} and \eqref{eq:second-equation}  provide a close set of partial differential 
equations for $\Omega_v$ and $\Omega_p$ only after some reasonable functional dependence of the local velocity $v_L$ from 
the dynamical variables $\Omega_v(x)$ and $\Omega_p$ is given.
%Thus we need to rely on a specific model for vortex motion that connects the vortex velocity to the forces acting on it. 
We find it  useful to introduce a  ``friction functional'' $\mathcal{B}$ defined as
\begin{equation}
\label{eq:general-B}
v_L^x \,\, = \,\, x \,\, \mathcal{B}[\Omega_v,\Omega_p,x] \,\, (\Omega_v(x)-\Omega_p) \,\,. 
\end{equation}
The explicit form of $\mathcal{B}$ depends on how vortex dynamics is modeled;
here we refer to a simple but widely used picture, where
a viscous force acts to deviate the vortices from corotation with the bulk superfluid
and this induces a hydrodynamical lift force (the Magnus force) on the vortices.

As suggested in \cite{EB92} a vortex line experiences a viscous drag force 
per unit length $\bm{f}_D$ and a Magnus force $\bm{f}_M$ expressed by
\begin{align*}
\bm{f}_D &= - \eta(\bm{x}) ( \bm{v}_L(\bm{x})-\bm{v}_p(\bm{x}) ) \, ,
\\
\bm{f}_M &=\rho_n(\bm{x}) \bm{\kappa} \times ( \bm{v}_L(\bm{x})-\bm{v}_n(\bm{x}) ) \, .
\end{align*}
As discussed in \cite{AS06}, the presence of entrainment does not modify the 
form of the Magnus force, providing that the vector $\bm{\kappa}$ is given by
\[
\bm{\kappa} \,=\, \kappa \,\nabla \times\bm{p_n} \, / \, |\nabla \times \bm{p_n}| \, .
\]
In the present case of straight vortices this vector is obviously directed along $\hat{\bm{e}}_z$.
Finally, the Kelvin's circulation theorem for ideal and barotropic fluids implies that 
free vortices in a background flow are transported with the fluid, thus behaving like massless 
particles.
Since there is no inertia, the equation of motion for straight and rigid vortices is given by
\[
\bm{F}_{tot} \,=\, \int_L \! dl(\bm{x}) \,[ \bm{f}_M(\bm{x})+\bm{f}_D(\bm{x}) ] \,=\,0 \, ,
\]
where the integration is performed over the line, as in Eqs \eqref{eq:a}-\eqref{eq:d}. In components the above equation reads
\begin{align*}
\int_0^{z(x)} \! dz\,& \left[ \eta(r)v_L^x(x) + \kappa \rho_n(r) (  v_L^{\theta}(x) - x \Omega_n(x,z) ) \right]=0
\\
\int_0^{z(x)} \! dz\,& \left[ \eta(r) (  v_L^{\theta}(x) - x \Omega_p ) -  \kappa \rho_n(r) v_L^x(x)  \right]=0 \, .
\end{align*}
Replacing the variable $\Omega_n$ with the aid of Eq. \eqref{eq:omegan} and using the definitions given in Eqs \eqref{eq:a}-\eqref{eq:d},
the solution of the system is
\begin{align}
\label{eq:line-velocity-solutions-rad}
\dfrac{v_L^x}{x} & \,=\,\dfrac{d \, b}{ d^2 + a^2} \,\, \omega
\\
\label{eq:line-velocity-solutions-theta}
\dfrac{v_L^\theta}{x} &\,=\,\dfrac{a \, b}{ d^2 + a^2} \,\, \omega \,+\,\Omega_p 
\end{align}
where the $x$-dependence is understood for notational simplicity. 
Here we have defined the angular velocity lag $\omega$ as
\begin{equation}
\label{eq:lag-definition}
\omega(x,t) = \Omega_v(x,t)-\Omega_p(t) \,\, .
\end{equation}
In the absence of entrainment our lag $\omega$ reduces to $\Omega_n-\Omega_p$.
Of course if $\Omega_v(x) \rightarrow \Omega_p$, vortices corotate locally with
the two components and there is no outward motion: $v^x_L(x) \rightarrow 0$ and $v^\theta_L(x) \rightarrow x \Omega_p$.
Equations \eqref{eq:line-velocity-solutions-rad} and \eqref{eq:line-velocity-solutions-theta} 
suggest to introduce the dimensionless parameters
\begin{align}
\label{eq:b-rad}
\mathcal{B}^x (x) \,&=\, \dfrac{d(x) \, b(x)}{ d(x)^2 + a(x)^2} \, \, ,
\\
\label{eq:b-theta}
\mathcal{B}^\theta (x) \,&=\, \dfrac{a(x) \, b(x)}{ d(x)^2 + a(x)^2} \,\, . 
\end{align}
Comparing Eq. \eqref{eq:general-B} and \eqref{eq:b-rad} 
we see that in the present approach the friction functional is only dependent on $x$, namely 
$\mathcal{B}[\Omega_v,\Omega_p,x] = \mathcal{B}^x (x)$.  
In the frame of reference of the crust, the velocity of vortex lines at radius $x$ 
makes an angle $\theta_d$ with the direction $\hat{\bm{e}}_\theta$ (the dissipation angle) such
that locally we have $\tan( \theta_d )= d(x)/a(x)$. This generalizes the
early result of  \citet{EB92} in the presence of entrainment and for a non-uniform star.
\subsection{Effective pinning}
\label{susec:effective-pinning}
Vortex pinning to the crustal lattice is the fundamental mechanism to store the angular momentum needed to produce a glitch.
One possibility to model pinning is to incorporate its effect in the general form of the functional $\mathcal{B}$,
in order to reproduce the local mean behavior of vortex lines. 
In Eq. \eqref{eq:line-velocity-solutions-rad} the
mean outward velocity $v_L^x$ is proportional to $\mathcal{B}^x(x)$ and pinning can be addressed 
by saying that only a fraction $Y$ of (unpinned) vortex lines is free to move under the action of the drag force. 
Formally this means that we are approximating the friction functional as
\begin{equation}
\label{eq:vel-y-b-w}
\mathcal{B}[\Omega_v,\Omega_p,x]\,\, \approx \,\,Y[\omega , x]\, \,\mathcal{B}^x(x) \, .
\end{equation}
Alternatively, by considering a large population of vortices around the radius $x$, 
we can also interpret $Y$ as the ``probability for unpinning'' \citep{JM06}.
To give an estimate of $Y$ we can use the general scenario of the snowplow mechanism described in \citet{P11}:
when the local lag $\omega(x)$ is greater than a critical lag for unpinning $\omega_{cr}(x)$, the vortices break free and are able to move 
outward. For clarity we stress that an important feature of the original snowplow model is that the particular form of $\omega_{cr}$ 
and the assumption of rapid expulsion of unpinned vortex lines
suggest the formation of a thin sheet of accumulated vorticity that slowly creeps outward. 
However, since our model is dynamical, we don't need to incorporate this ``vortex-sheet'' assumption:
here we just need to generalize the prescription for calculating the snowplow critical profile $\omega_{cr}$ when entrainment is present.
%For clarity we stress that our aim is to incorporate the general snowplow picture into our hydrodynamical description
%but without explicitly using the vortex-sheet hypothesis; here we just need to generalize 
%the prescription for calculating the snowplow critical profile $\omega_{cr}$ when entrainment is present.

The starting point is the mesoscopic vortex-lattice interaction, namely the {\em average} pinning force (per unit length) acting on a rigid vortex that 
passes trough many lattice sites.
\citet{DP04,DP06} studied the microscopic vortex-nucleus interaction, whereas 
\citet{SP15} provide the mesoscopic pinning force per unit length in the crust $f_P(\rho)$. 
The physical meaning of this quantity is that
a segment of vortex line can locally unpin if the modulus of the forces (leaving out the pinning force) acting on it equals $f_P$.
In the following we present a simple extension of this to the rigid vortex motion described here,
based on a ``rigid'' model of the pinning force: a  vortex line unpins when the modulus of the total force acting on it 
is equal or greater than the integrated value of $f_P$ along the line.

\emph{Rigid pinning} - When a pinning force per unit length 
$\bm{f}_P(\bm{x})$ acts on a segment of vortex line at $\bm{x}$ ,  the general equations of motion of rigid vortices become
\[
\bm{F}_{tot} \,=\, \int_L \! dl(\bm{x}) \,[ \bm{f}_M(\bm{x})+\bm{f}_D(\bm{x})+\bm{f}_P(\bm{x}) ] \,=\,0 \, . 
\]
The vectorial quantity $\bm{f}_P(\bm{x})$ is not known, since no reliable estimate of the dependence of the microscopic pinning force on the vortex-nucleus separation is available. Note that, despite our notation, in general $f_P$ is not the modulus of $\bm{f}_P$: the mesoscopic quantity $f_P(\rho)$ studied by \cite{SP15} represents the threshold for unpinning, namely it is the maximum value that the pinning force can sustain before letting the vortex segment free to move (analogous to a static friction force, which adjust itself to the external force up to a maximum value where motion sets in).
Despite this difficulty, we can use the equation of motion to determine the critical lag for unpinning; the viscous drag force is such that $|\bm{f}_D|_{\bm{v}_L =\bm{v}_p} =  0$, namely it does not act on pinned vortices comoving with the crust. Thus
we can write the condition for unpinning in terms of the Magnus force alone, whatever the actual form of $\bm{f}_P$ is:
due to the macroscopic rigidity of the vortex line, 
the Magnus force acting on the line (at a cylindrical radius $x$) must be greater than 
\[
F_P(x) \, = \,2 \int_0^{z(x)} f_P(r)\,\,dz 
\]
in order to unpin the vortex. It follows that
\[
F_P(x) \, = \, |\bm{F}_M(x)|_{\bm{v}_L =\bm{v}_p} 
\]
is the magnitude of the total Magnus force acting on a vortex when the critical velocity lag for unpinning is reached. 
Since in this case
\[
|\bm{F}_M(x)|_{\bm{v}_L =\bm{v}_p}  =  2 \int_0^{z(x)} \! dz \,
\frac{\kappa\,\rho_n(r)}{1-\epsilon_n(r)} \left( v_p(x) - \frac{p_n(x)}{m_n} \right) \, ,
\]
by using the definition of lag given in Eq. \eqref{eq:lag-definition} and the definition of $b(x)$ given in Eq. \eqref{eq:b}
it is straightforward to find that the critical lag profile is given by 
\begin{equation}
\label{eq:cr-lag}
\omega_{cr}(x) = \dfrac{ F_P(x) }{ \kappa \, x \, b(x) } \, \, .
\end{equation}
With no entrainment  ($\epsilon_n=0$) this result reduces to the definition of the critical lag profile given in \cite{P11}.
%The  appearance of $b(x)$ at the denominator is physically reasonable: the effect of entrainment 
%(in the frame of reference of the crust) is to redefine the mass of neutrons as $m^*_n/m_n= 1-\epsilon_n$ that rescales $\rho_n$.

The simplest way to encode pinning in our equations is thus to choose 
\begin{equation}
\label{Y-step}
Y[\omega , x]=\theta(|\omega(x)|-\omega_{cr}(x))   \, .
\end{equation}
When the absolute value of the local lag is smaller than $\omega_{cr}(x)$ the fraction of unpinned vortices is zero
and there is no radial motion of vortex lines; otherwise the vortices contained in a thin cylindrical shell
of radius $x$ can move inward or outward depending on the sign of $\omega(x)$.
The introduction of such a non analytical term into the equations poses a mathematical complication. However we have in mind
that our equations can be used to define a numerical simulation of pulsar glitches: discretizing the cylindrical radius 
in many cells, Eq. \eqref{Y-step} just tells us whether vortices in that cell can be moved or not.

\emph{Vortex-creep} - By using Eq. \eqref{Y-step} we have the advantage that there is no need to introduce new parameters into the model,
even though it is an unfortunate choice from the analytical point of view.
We expect that temperature, quantum effects and finite vortex tension all take part in smoothing the step-like form of $Y$, 
somewhat similarly to the early vortex-creep model of \cite{AA84} where the radial velocity is given by
\[
%v_L^x \, \approx \, v_0 \, \exp \left(  \frac{\omega(x) - \omega_{cr}(x)}{\alpha(x) \,\omega_{cr}(x)}  \right) \, .
v_L^x \, \approx \, 2 \,v_0  \,e^{ -1/\alpha(x) } \sinh{\left( \frac{\omega(x)}{\alpha(x) \, \tilde{\omega}_{cr}(x)} \right)}  \, .
\]
%thermally-activated vortex creep velocity %
Here $\alpha$ is a dimensionless temperature that tunes the local rigidity of the pinning, $v_0$ a typical velocity 
of the microscopic motion of vortex lines and $\tilde{\omega}_{cr}(x)$ is an opportune critical lag. 
In the framework of the original work of \cite{AA84} it is not simple to implement consistently the geometry of the problem 
(even with our strong assumption of straight and rigid vortices) and it is not clear which prescription we have to use to estimate 
the cylindrical profiles $\alpha(x)$ and $\tilde{\omega}_{cr}(x)$.
However the expression for $v_L^x$ of \cite{AA84} physically motivates the role of temperature and tells us that we can study the effect 
of a less rigid pinning by using a smooth $Y$ functional;
%It is thus tempting to use the above expression to envisage how we can study the effect of a less rigid pinning in our model: 
%we can introduce a $Y$ functional such that in the limit $\alpha \rightarrow 0$ we recover Eq. \eqref{Y-step};
for example one of the simplest possibilities is
\[
Y[\omega , x] = \frac{1}{2} + \frac{1}{2} \tanh{\left( \frac{|\omega(x)|-\omega_{cr}(x)}{\alpha \, \omega_{cr}(x)} \right)}  \, ,
\] 
where in the limit of ``zero temperature and rigid vortices'' $\alpha \rightarrow 0$ we recover Eq. \eqref{Y-step}.
The disadvantage with respect to the original approach of \cite{AA84} is that now $\alpha$ has no direct physical interpretation.

\emph{Viscous drag} - Another way to encode pinning that is different from Eq. \eqref{eq:vel-y-b-w} is to consider a large value of 
the drag parameter $\eta$ in the crust: suppose that $\eta \rightarrow \infty$ in the interval $R_c<r<R_d$, 
then $d(x) \rightarrow \infty$ for every value of $x$. 
%Another possibility to encode pinning that is different from Eq. \eqref{eq:vel-y-b-w} is to consider a large value of 
%the drag parameter $\eta$ in the crust: suppose that $\eta \rightarrow \infty$ in some region of the interval $R_c<r<R_d$, 
%then $d(x) \rightarrow \infty$ for every value of $x<R_d$. 
By means of Eqs \eqref{eq:line-velocity-solutions-rad} and \eqref{eq:line-velocity-solutions-theta}, 
we have $v_L^x=0$ and $v_L^\theta=x\Omega_p$, namely in the whole star all the vortices are pinned and co-move with the p-component.
The idea is thus to use a viscous drag parameter $\eta[\rho,\omega]$ that has a local dependence on the lag,
so that $\eta$ assumes a very high value when $|\omega(x)|$ is well below $\omega_{cr}$.
It is not surprising that a strong viscous drag is equivalent to pinning, but the practical problem is that for every 
value of $\omega(x)$ we have to calculate $\mathcal{B}^x(x)$ using the opportune values $\eta[\omega(x),\rho(r)]$.
%, where $r$ varies while keeping its projection on the equatorial plane $x$ fixed. 
%
%A strong viscous drag is thus equivalent to pinning and 
%we can use as dynamical friction functional the $\mathcal{B}^x(x)$ of Eq. \eqref{eq:b-rad} providing that the integration of $\eta[x,\omega(x)]$ gives 
%$d(x) \gg a(x)$ for $|\omega(x)|<\omega_{cr}(x)$. 
% NOTA: QUI SOTTO VA BENE SIA ESPONENTE 21 CHE 22... STIMA MOLTO ROZZA...
As an order of magnitude estimate, $\eta \gtrsim 10^{21}\,$s cm g$^{-1}$ in the crust 
is sufficient to give a friction parameter ranging from $\mathcal{B}^x(0) \sim 10^{-9}$ to $\mathcal{B}^x(R_d) \sim 10^{-12}$; this  ensures that the vortices are almost blocked on timescales of many years (cf. equation \eqref{eq:relax-time}). On the other hand, a rapid expulsion of vortex lines occurs when $\eta$ is such that $d(x)\approx a(x)$ and the dissipation angle is nearly $\pi/4$.
\section{Macroscopic equations for the two fluid model and relation with a rigid model}
\label{sec:macro-eqs}
In the previous section we derived the macroscopic equations of our cylindrical model, Eqs \eqref{eq:first-eq-gen} and \eqref{eq:second-equation},
where the physical (observable) quantity is $\Omega_p$.
With the aid of the definition given in Eq. \eqref{eq:radial-mean} we can rewrite this system for
$\Omega_v(x,t)$ and $\Omega_p(t)$ in a more compact form as
\begin{align}
\label{eq:final-sys-v}
    \partial_t{\Omega}_v&(x,t) =
    - \mathcal{B}[\Omega_v,\Omega_p,x]\,(2\Omega_v + x\partial_x \Omega_v)(\Omega_v-\Omega_p)
    \\
    \label{eq:final-sys-p}
    \partial_t{\Omega}_p&(t) = 
    -(1+q)\dot{\Omega}_\infty - q \langle \, \partial_t {\Omega}_v(t) \, \rangle 
\end{align}
where $q=I_v / I_p$ and $\dot{\Omega}_\infty = T_{ext}/I$, while the definition of the cylindrical mean, weighted by $dI_v(x)/I_v=g(x) dx$,  
is given in Eq. \eqref{eq:radial-mean}.
The first remark is that the presence of entrainment does not modify the form of the above equations: all the effects of entrainment are contained in the renormalized quantities $q$ and $\mathcal{B}$, provided one uses as effective rotational variables  $\Omega_v(x,t)$ and $\Omega_p(t)$, which are associated to the effective moments of inertia $I_v$ and $I_p$ respectively; in this sense we can interpret the v-component, directly related to the vortex density, as the actual reservoir of angular momentum when entrainment is present. Also, we can trigger a glitch by perturbing the $\mathcal{B}$ functional (or by suddenly increasing the $Y$ fraction in some interval of the variable $x$):
this mimics the sudden unpinning of many vortices.

%with our formalism 
%the entrainment only modifies the specific values of $q$, $\mathcal{B}$ and of the weight $g$.
It is worth to note that if in Eq. \eqref{eq:final-sys-p} we constrain $\partial_t \Omega_p=\partial_t \Omega_v$ for all the $x$ values, it
follows that the star responds to the braking torque as a whole, with a spin-down rate of $-\dot{\Omega}_\infty$. 
As usual, we refer to this situation as ``steady-state''; the value of $\dot{\Omega}_\infty$ can be settled by observing 
the mean spin-down of the pulsar over a long period of time and of course is independent of entrainment. 
%We stress that this is a definition 
%of a possible dynamical state of our theoretical model
%and in principle we do not know if is it possible for a real pulsar to spin down as a whole
%(for example, there could be not enough time between two consecutive glitches to reach this dynamical state).

As described in the previous section, the non uniform structure of the star and the presence of entrainment influence
the friction functional $\mathcal{B}$, the ratio $q$ and in particular the cylindrical mean.
On the other hand, the actual value of the stellar radius plays no role since the above equations do not change under 
the rescaling $x \rightarrow x  R_d$ and $g \rightarrow g / R_d $.  In the following we will thus measure the radius in units of $R_d$, namely $x$ lies in the range $0\leq x \leq1$.

As a direct consequence of the definition of $g$, we can calculate the angular momentum of the differential v-component as
\[
L_v(t) \, = \, I_v \, \langle \Omega_v(x,t)  \rangle \, .
\]
It follows that, given a generic lag profile $\omega(x)$, the reservoir of angular momentum associated to the lag is
\begin{equation}
\label{eq:ang-mom-reservoir}
\Delta L[\omega] \,=\, L_v  \,-\,  I_v\, \Omega_p       \,=\, I_v \, \langle \omega(x) \rangle \, .
\end{equation} 
In the last section, we will describe how this fact can be used to set an upper limit to the mass of a large glitcher like the Vela.
%By assuming that a real pulsar with observed angular velocity $\Omega_{ob}$
%is in the steady state when it
%displays no changes in the observed spin down 
%rate $\dot{\Omega}_{ob}$ over timescales comparable with the typical inter-glitch waiting time, we can estimate an upper bound to 
%the angular momentum reservoir stored in the superfluid component for that pulsar:
%\begin{equation}
%\label{eq:steady-reservoir}
%\Delta L_\infty \,=\, I_v \, \langle \omega_\infty \rangle \, \, ,
%\end{equation}
%where the steady-state lag $\omega_\infty$ is time independent and can be found by searching for a solution of the non linear equation 
%\begin{equation}
%\label{eq:steady-lag-general}
%\mathcal{B}[\Omega_{ob}+\omega_\infty,\Omega_{ob},x] \,
%(2\Omega_{ob}+2\omega_\infty+x\partial_x \omega_\infty)\, \omega_\infty \,= \,\dot{\Omega}_{ob}
%\end{equation}
%with a suitable boundary condition. 
%RIVEDERE QUESTA MINI SPIEGAZIONE...\\
%Equations \eqref{eq:final-sys-p} and \eqref{eq:final-sys-v} tell us that the angular momentum reservoir $\Delta L$ is maximal at steady state
%since an hypothetical lag greater than the steady state lag $\omega_\infty$ must relax rapidly: 
%if we let the system evolve starting from an initial lag $\omega_0$ such that $\langle\omega_0 \rangle > \langle\omega_\infty\rangle$,
%we have that $\Omega_p(t)>\Omega_p(0)-t \dot{\Omega}_\infty $ and $\langle \partial_t{\omega}\rangle <0$.
%
\subsection{Reduction to a rigid model}
\label{subsc:reduction-rigid-model}
Most studies of pulsar glitches are based on  body-averaged 
models with two rigid components: an ``observable component'' strongly coupled to the magnetic field 
and an ``internal component''. Some examples of this kind of approach can be found in the early work 
of \cite{BP69} or in the more recent studies of \cite{AG12}, \cite{SP10} and \cite{J02}.
The form of the mutual friction is assumed to be linear in the velocity lag and the dynamical equations are 
\begin{align}
\label{eq:baym1}
& I_o \dot{\Omega}_o + I_i \dot{\Omega}_i = - I \dot{\Omega}_\infty
\\
\label{eq:baym2}
& \dot{\Omega}_i =  - \frac{I_o}{I} \,\frac{\Omega_i - \Omega_o}{\tau} \, ,
\end{align}
where the label $i$ ($o$) refer to the internal (observable) component 
and $\tau$  is the phenomenological timescale that describes the coupling between the components. %(the ``relaxation time'').

In the original work of \cite{BP69}, the aim was to study the response induced by a sudden change in the moments of inertia, thus the two components
are the crust (plus the charged particles) and the totality of the superfluid neutrons; this gives $I_o \sim 10^{-2} I$.
On the other hand, under the assumption that glitches arise in the inner
crust superfluid (taken as distinct from the core superfluid), the observable component is comprised by the normal crust plus the totality of the core. 
In this case the core 
turns out to be strongly coupled to the normal component thanks to the effect of electron scattering off magnetized vortex cores \citep{AL84},
another manifestation of entrainment;
thus estimates of the moment of inertia of the superfluid in the crust give $I_i \sim 10^{-2} I$ \citep{LE99}, 
which is a quite small reservoir of angular momentum \citep{P11}. 

As argued by \cite{AG12} and \cite{C13}, the fraction $I_i/I$ is significantly
lowered by  entrainment, leading to serious difficulties in explaining large glitches.
%The snowplow model is a viable possibility to overcome this problem, but at the moment there is no any
%detailed study of the effect of entrainment on this model. 
For this reason we decided to work under the assumption of vortex cores that pass trough the core-crust interface (as in the snowplow model),
since this gives a larger angular momentum reservoir. 
It is comforting that this choice is so far favored microscopically.
%This point will be addressed more extensively the last section of this work, together with numerical estimates of the angular momentum reservoir. 

It's  interesting to study under which conditions the rigid model defined by Eqs \eqref{eq:baym1} and \eqref{eq:baym2}
can be a reasonable approximation of the Eqs. \eqref{eq:final-sys-v} and \eqref{eq:final-sys-p}.
For this purpose we define the deviation $\delta f$ of a generic function $f$ as
\[
f(x,t) = \delta f (x,t) + \langle f(t) \rangle
\] 
and an ``angular acceleration'' quantity $\mathcal{A}(x,t)$ as
\[
\mathcal{A} = ( \langle\mathcal{B}\rangle + \delta \mathcal{B} ) 
(2 \Omega_p + 2 \langle \omega \rangle + 2 \delta \omega + x \partial_x \delta \omega) 
(\langle \omega \rangle + \delta \omega) \, .
\]
We can rewrite the Eqs \eqref{eq:final-sys-v} and  \eqref{eq:final-sys-p} as 
a system of three dynamical equations for two rigid variables 
(the mean lag $\langle \omega \rangle$ and $\Omega_p$, that depend only on time) and a differential one $\delta \omega(x,t)$:
\begin{align}
\label{eq:sys1}
& \delta \dot{\omega} = -\mathcal{A} + \langle \mathcal{A} \rangle
\\
\label{eq:sys2}
& \langle \dot{\omega} \rangle = (1+q) ( \dot{\Omega}_\infty - \langle \mathcal{A} \rangle )
\\ 
\label{eq:sys3}
& \dot{\Omega}_p = - (1+q) \dot{\Omega}_\infty + q \langle \mathcal{A} \rangle \, .
\end{align}
It's possible to approximate $\mathcal{A}$ in such a way that 
the Eqs \eqref{eq:sys2} and \eqref{eq:sys3} decouple from Eq. \eqref{eq:sys1}.
A very general $\mathcal{B}$ functional makes the problem analytically difficult. On the other hand we can restrict 
our attention to a finite time interval during which we assume that the friction functional $\mathcal{B}$ does not vary rapidly over time
and can be approximated with an effective cylindrical profile that is function of $x$ only, namely $\mathcal{B}\approx \mathcal{B}^x_{\text{eff}}(x)$, 
where $\mathcal{B}^x_{\text{eff}}$ eventually accounts also for the fraction $Y$.
%Since $\langle \delta f \rangle =0$ by definition, it's clear that, namely
%\[
%mathcal{A} \approx 2 (\langle\mathcal{B}^x\rangle + \delta \mathcal{B}^x) \Omega_p \langle \omega \rangle 
%+2 \langle\mathcal{B}^x\rangle \Omega_p \delta \omega
%\]
%in such a way we find the original system proposed by Baym for two rigid components with angular velocities
With this assumption, 
we can decouple  equation  \eqref{eq:sys1} from equations \eqref{eq:sys2} and \eqref{eq:sys3} 
by dropping the term $x \partial_x \delta \omega$ and taking in $\mathcal{A}$ only the terms that are up to first order in the deviations:
\begin{equation}
\label{eq:rigid-approx}
\langle \mathcal{A} \rangle \approx 2 \langle \mathcal{B}^x_{\text{eff}} \rangle ( \Omega_p  +\langle \omega \rangle)
\langle \omega \rangle  \approx
2 \langle \mathcal{B}^x_{\text{eff}} \rangle \, \Omega_p \, \langle \omega \rangle \, .
\end{equation}
In this way the dynamics of $\Omega_p$ and $\langle \omega \rangle$ does not depend on $\delta \omega$.
%and we can restrict our attention to the two rigid equations only. 
Now, if we identify $\Omega_o$ with $\Omega_p$ and $\Omega_i$ with $\Omega_p+\langle \omega \rangle$, the
Eqs \eqref{eq:sys2} and \eqref{eq:sys3} are equivalent to Eqs \eqref{eq:baym1} and \eqref{eq:baym2},
where $q=I_i/I_o$. We will see later that $q \approx 10-20$ for realistic star parameters; thence we have a potentially large reservoir of angular momentum.
It also follows that the relaxation time can be estimated as 
\begin{equation}
\label{eq:relax-time}
\tau ^{-1}\, =\, 2  (1+q) \langle \mathcal{B}^x_{\text{eff}} \rangle \Omega_p \, .
\end{equation}
The relaxation time given here should be intended as the coupling timescale of a body-averaged model characterized by
$\langle \mathcal{B}^x_{\text{eff}} \rangle$. 
In this way, thanks to our cylindrical mean, we can give a microphysical interpretation of the phenomenological parameter $\tau$.
%Obviously the differential equations \eqref{eq:sys2} and \eqref{eq:sys3} give rise to relaxation solutions that are not exponential,
%thus 
%It is our opinion that the differential and non linear terms of our equations
%are important mainly because of their effect on driving the functional $\mathcal{B}[\Omega_v,\Omega_p,x]$.
We have also shown that the entrainment effect does not change the form of the rigid equations but only affects the values 
of $I_o$, $I_i$ (leaving $I$ unchanged) and $\tau$, without any need to introduce new terms. 
\section{Numerical integration of the input profiles}
\label{sec:num-integration}
To use the dynamical equations \eqref{eq:final-sys-v} and \eqref{eq:final-sys-p} we first need to set
the ratio $q$ and the normalized function $g(x)$. This will fix the structure of the star.
These parameters depend on the mass of the star and on the EOS of dense matter, as well as on entrainment. 
On the other hand, all the information about the dynamics of vortices is contained in $\mathcal{B}$;
if the prescription of Eq. \eqref{eq:vel-y-b-w} is used, we have to calculate a static drag-parameter profile $\mathcal{B}^x(x)$
and the critical profile $\omega_{cr}(x)$.
These three functions of $x$ and the number $q$ are the macroscopic ``input profiles'' of the model and can be obtained from micro and mesoscopic physics
using the integrations described in the previous sections; to construct them we need the density profile
$\rho(r)$ and the density dependent quantities $x_n(\rho)$, $\epsilon_n(\rho)$, $\eta(\rho)$ and $f_p(\rho)$.

Since in this preliminary work we are dealing with Newtonian moments of inertia,
in order to keep the exposition simple we do not implement the general relativistic conversion between the baryon
number density $n$ and the energy density $\rho$, namely  we take $n = \rho / m_n $ and $n_n = x_n \rho / m_n  $.
We will study this aspect, by incorporating the  prescriptions of \cite{HP04}
for a thermodynamically consistent EOS as well as the corrections for slow rotation in general relativity, in a dedicated future work.
%To re-parametrize the profiles expressed in terms of the baryon number
%density as functions of the mass baryon density we use the prescriptions described in \citep{HP04}
%for a thermodynamically consistent EOS. 
\subsection{Microphysical input}
For a given central density and EOS we solve the TOV equations. This gives the total mass of the neutron star and the
density profile $\rho(r)$ so that we can express all the input profiles as functions of the spherical radius.

%
%EOS AND XN%EOS AND XN%EOS AND XN%EOS AND XN%EOS AND XN%EOS AND XN%EOS AND XN%EOS AND XN%EOS AND XN%EOS AND XN
%
\begin{figure}
	\centering
   	\includegraphics[width=.45 \textwidth]{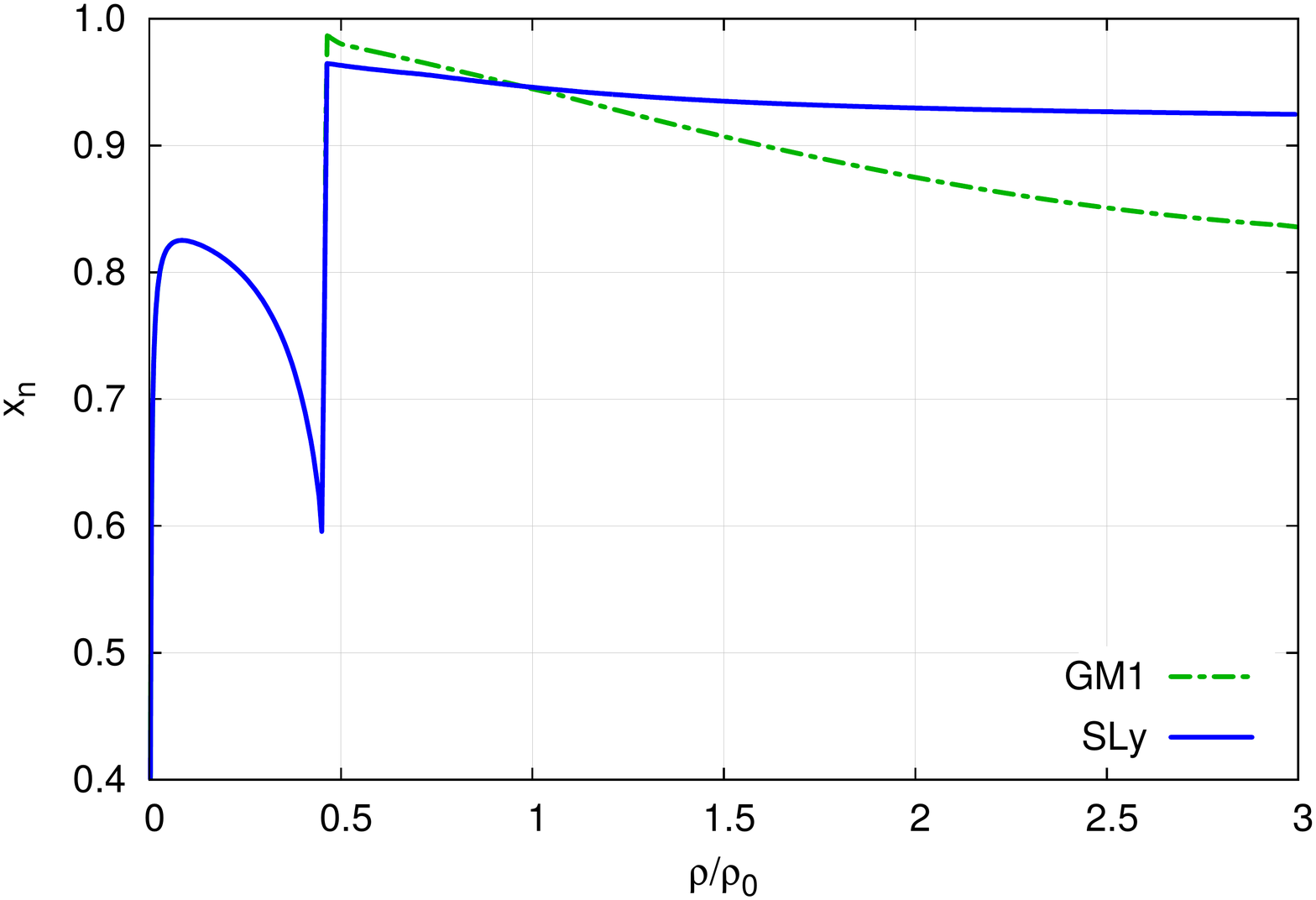}
	\caption{The superfluid fraction $x_n$ as a function of the baryon density used in our examples: the blue solid line is consistent with the Sly EOS,
	obtained by joining the data in Table 1 and Table 4 of \citet{DH01}. 
	Here we show also the superfluid fraction consistent with the GM1 EOS in the core (dashed line).
	For densities below the drip point we have $x_n=0$ (no superfluid neutrons are present). 
    In the inner-crust $x_n$ is the fraction of superfluid nucleons in the neutron gas outside nuclei.
	}
	\label{fig:xn}
\end{figure}
\emph{EOS} - We use the SLy EOS proposed in \cite{DH01} as a consistent description of the whole star.
It provides the relation $P(\rho)$ and the fraction of free superfluid neutrons $x_n(\rho)$ (shown in Fig. \ref{fig:xn}). 
This equation of state predicts a first order phase transition  where nuclear clusters dissolve into homogeneous nuclear matter 
at baryon density $n \sim 0.76$ fm$^{-3}$, corresponding to $\rho_c \approx 1.3 \times 10^{14}$ g cm$^{-3}$. 
To find the value of $R_d$ we assume that neutron drip appears at density $\rho_d = 4.3 \times 10^{11}\,$g cm$^{−3}$. 

To study the dependence from the EOS we also use the stiffer GM1 equation of state of \cite{GM91} in the core, 
while keeping the SLy to describe the crust. The corresponding $x_n$ fraction is shown in Fig \ref{fig:xn}.
In all the figures we measure the density in units of $\rho_0 = 2.8\times 10^{14}\,$g cm$^{-3}$, the nuclear saturation density.

%
%ENTRAINMENT%ENTRAINMENT%ENTRAINMENT%ENTRAINMENT%ENTRAINMENT%ENTRAINMENT%ENTRAINMENT%ENTRAINMENT%ENTRAINMENT%ENTRAINMENT 
%
\begin{figure}
	\centering
    \includegraphics[width=.47 \textwidth]{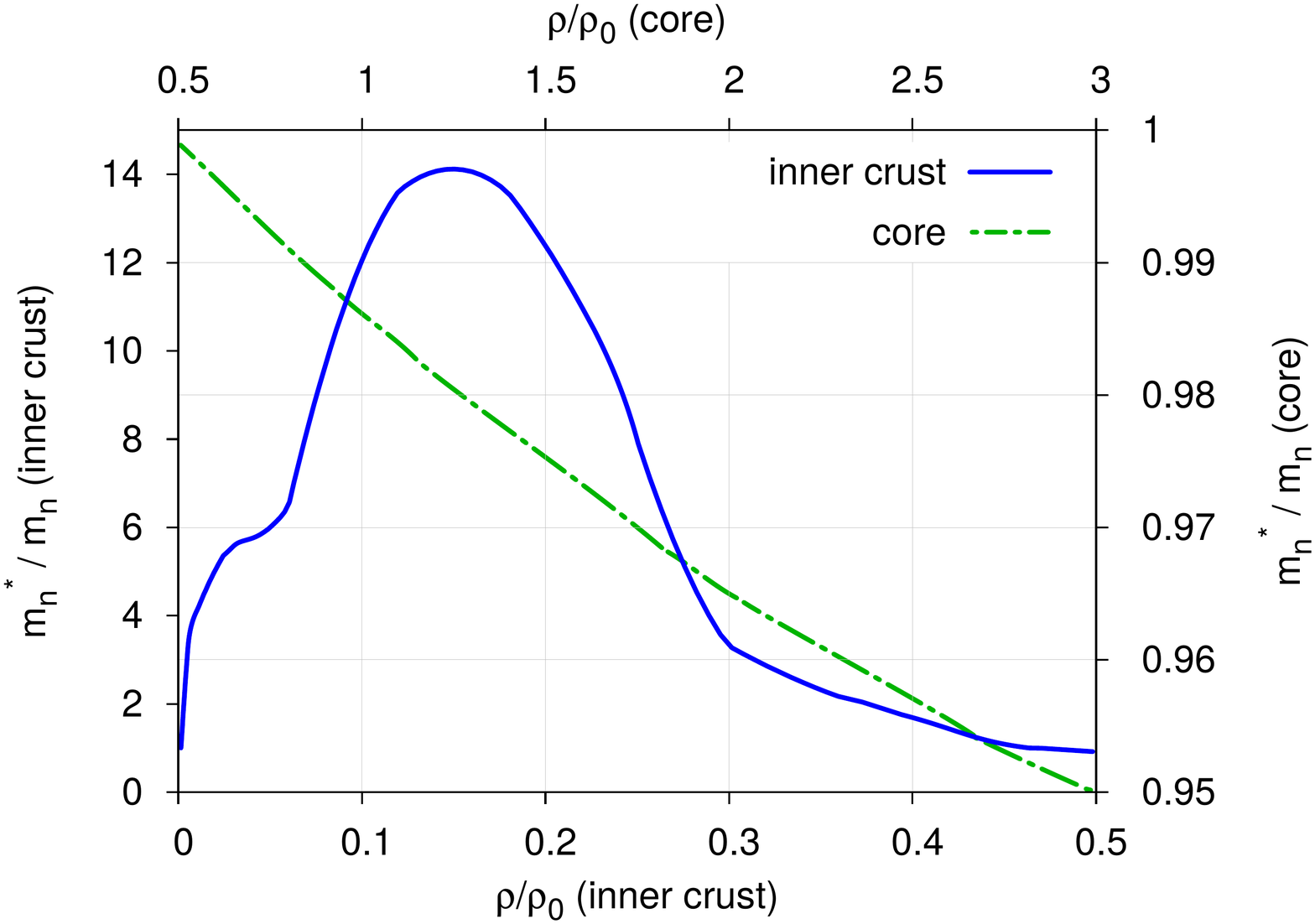}
	\caption{Effective neutron mass in the crust (obtained by interpolating the values given in \citet{C12}, solid line) 
    and in the core (from \citet{CH06}, dashed line). The crust-core interface is at $\rho_c \approx 0.47 \, \rho_0$. 
	}
	\label{fig:mn}
\end{figure}
\emph{Entrainment} - Superfluid entrainment in neutron stars was studied by Chamel and collaborators in a series of articles. 
Here we use the effective neutron mass $m_n^*(\rho)$
given in \cite{C12} and in \cite{CH06} for the crust and the core respectively, as shown in Fig. \ref{fig:mn}. 
The entrainment parameter is $\epsilon_n = 1-m_n^*/m_n$ where $m_n$ is the bare nucleon mass.  It is evident from the figure that $m_n^*$ depends strongly on density.
Even though the neutron superfluid 
can flow without friction in the crust, it is entrained by the nuclei because of Bragg scattering of neutrons by 
the crustal lattice \citep{CC05}. The entrainment parameters that we use in this work are easily obtained as 
$\epsilon_n = 1- m_n^*/m_n$ and $\epsilon_p= x_n \epsilon_n/(1-x_n)$. 
%At the transition between crust and core the effective mass is settled to unity.

%
%F PINNING%F PINNING%F PINNING%F PINNING%F PINNING%F PINNING%F PINNING%F PINNING%F PINNING%F PINNING%F PINNINGF PINNING
%
\begin{figure}
	\centering
	\includegraphics[width=.45 \textwidth]{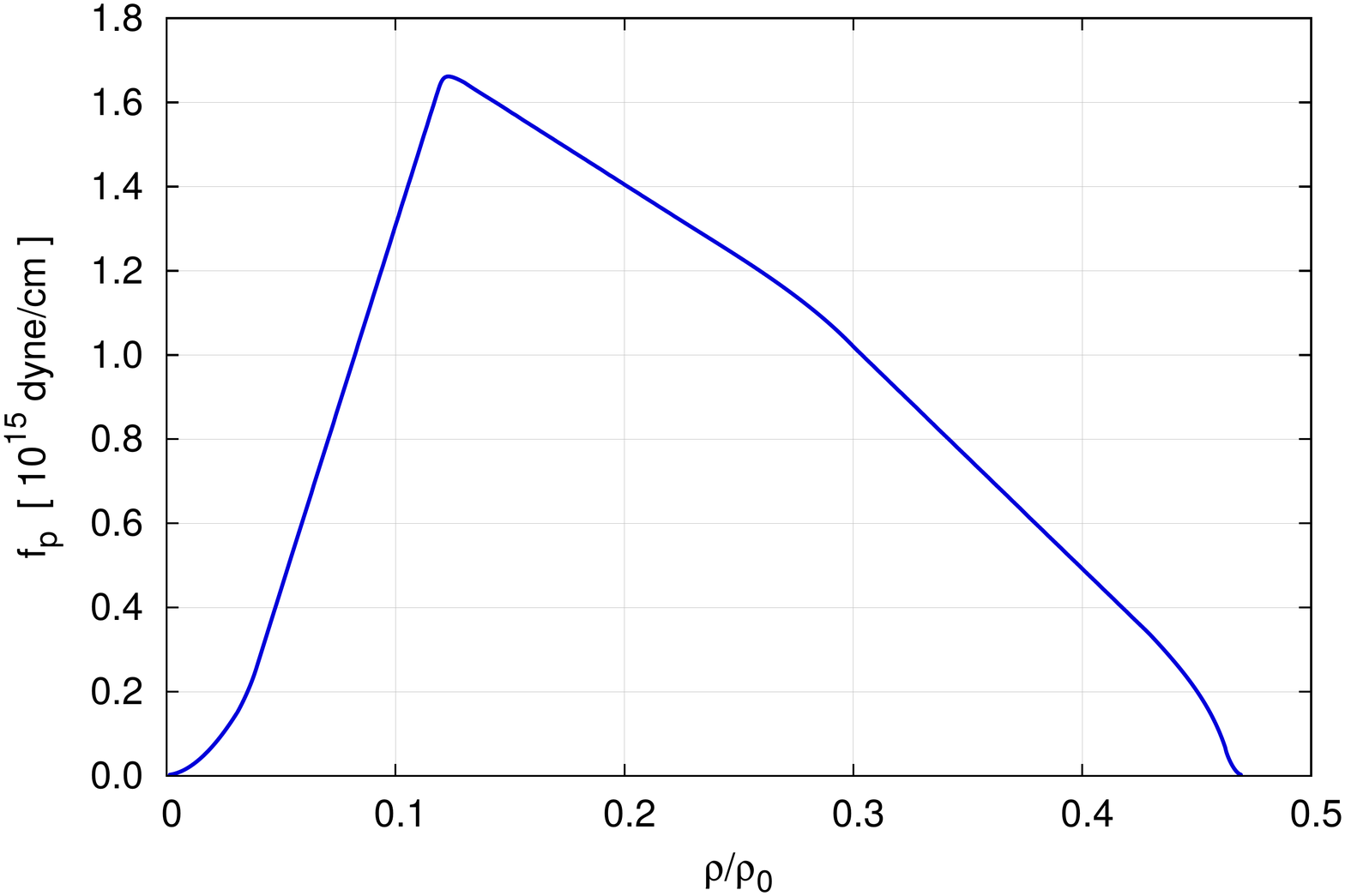}
	\caption{The pinning force profile per unit length $f_p$ used to calculate $\omega_{cr}$ in units of $10^{15}\,$dyne/cm.
             We constrain the pinning force profile to be zero outside the density range $\rho_d < \rho < \rho_c$.}
	\label{fig:fp}
\end{figure}
\emph{Pinning force} - Another important ingredient of the model is $f_p$,
the pinning force per unit length resulting from the vortex-lattice interaction in the inner crust.
Recently \citet{SP15} have proposed a numerical simulations to evaluate $f_p(\rho)$  at different densities in the inner crust,
which accounts for finite vortex tension and random orientation with respect to the lattice. 
In particular we use one of the pinning profile corresponding to in-medium suppressed pairing gap 
(the case $\beta=3$ and $L=5000$ of \cite{SP15}, see table 3 therein), as shown in Fig. \ref{fig:fp}.
Since theoretical estimates suggest that the protons in the core could
form a type II superconductor, in principle the vortex motion could also be impeded by the interaction with the flux tubes.
However, following the phenomenological study of \citet{HP13} this pinning interaction should be small (we are currently studying this issue along the lines of \cite{SP15}). For this reason we choose to constrain $f_p$ to be zero outside the crust.

%
%DRAG%DRAG%DRAG%DRAG%DRAG%DRAG%DRAG%DRAG%DRAG%DRAG%DRAG%DRAG%DRAG%DRAG%DRAG%DRAG%DRAG%DRAG%DRAG%DRAG%DRAG%DRAG%DRAG%DRAG
%
\begin{figure}
	\centering
	\includegraphics[width=.45 \textwidth]{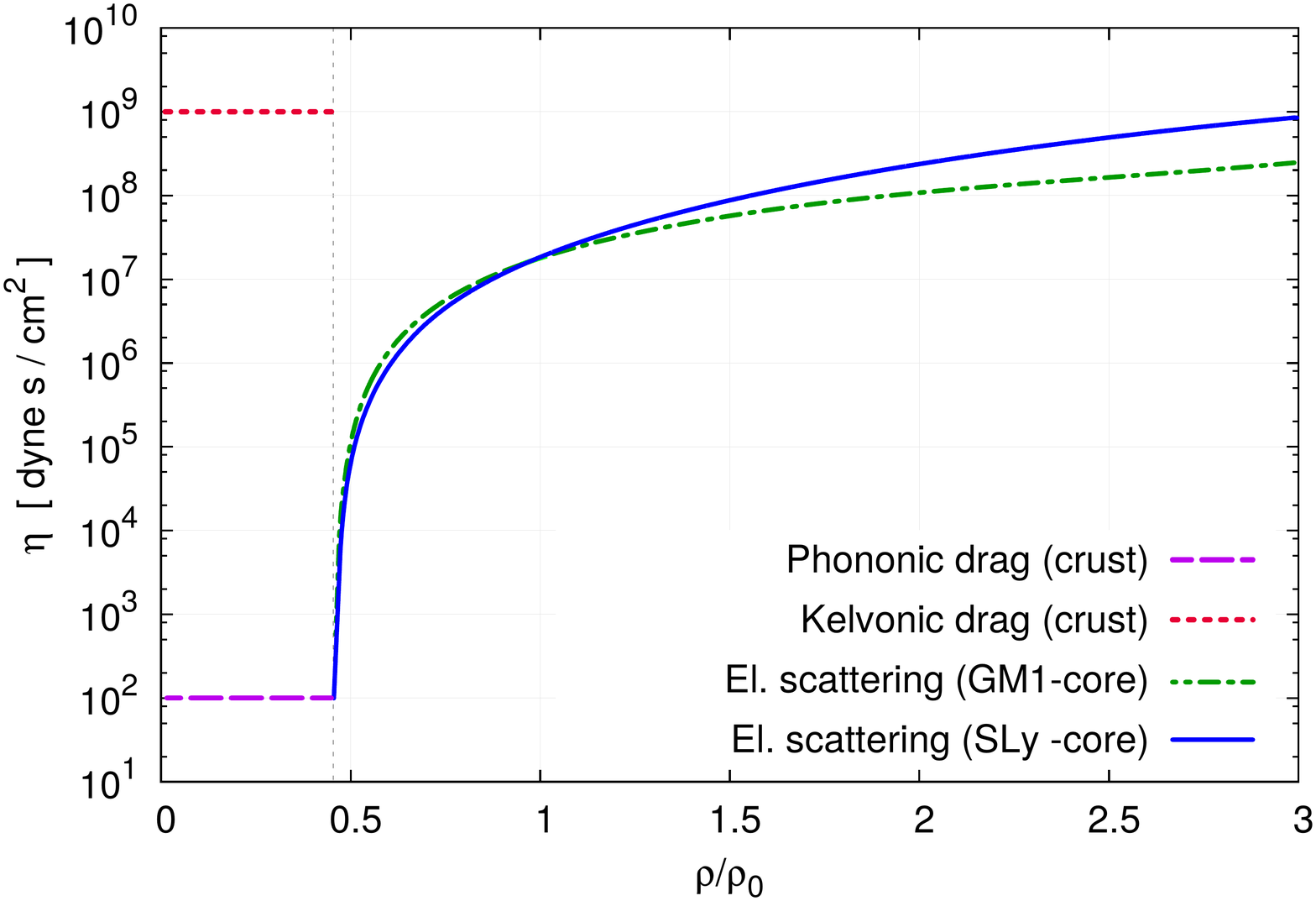}
	\caption{Example of viscous drag parameter $\eta(\rho)$: in the core ($\rho  > 0.47 \rho_0$) 
	the dominant dissipative channel is the electron scattering off magnetized vortex lines, thus $\eta(\rho)$ is given by Eq. \eqref{eq:drag-ander}.
	In the crust, for $\rho_d<\rho<\rho_c$, we consider two dissipative mechanisms: the excitation of phonons in the 
	crustal lattice (purple dashed line) or the excitation of Kelvin waves in the vortex cores (red dotted line).
	}
	\label{fig:eta}
\end{figure}
\emph{Viscous drag} - The drag parameter $\eta$ is, probably, the most difficult and less studied microphysical input that we need:
a realistic treatment of the drag exerted on vortices implies the study of many dissipative phenomena, although
we can take into account only the dominant ones in each region of the star.
\cite{AS06} give an estimate of the drag on vortex lines due to the electron scattering off 
magnetized cores as 
\begin{equation}
\label{eq:drag-ander}
\eta_{es} \approx  6.1 \times 10^{-5} \, \kappa  \,
x_n  \, x_p^{7/6} \, \rho^{7/6}  \frac{\epsilon_p^2}{\sqrt{1-\epsilon_p}} 
\left( \frac{\text{cm}^3}{\text{g}}\right)^{1/6}
\end{equation} 
where $x_p=1-x_n$, $\epsilon_p = x_n \epsilon_n/x_p$ \citep{CH06} and all the quantities are functions of density.
The standard picture is that vortices in the core are magnetized, due to the entrainment effect 
between the neutron superfluid and the proton superconductor. For this reason the above estimate is
valid only for $\rho > \rho_c$, thus we take $\eta_{es}=0$ in the crust.

On the other hand, as pointed out by \citet{J90a} and \citet{EB92}, for $\rho < \rho_c$ the vortices 
excite phonons in the crustal lattice. This effect dominates the dissipation when the relative velocity between
the vortex line and the lattice is small ($|\bm{v}_L-\bm{v}_p |<10^2$ cm s$^{-1}$).
%\citet{J90a} estimates the viscous drag parameter due to the excitation of lattice phonons as
%
%\[
%\eta_{ph} \,  \approx \, 0.053\,\frac{a}{\xi^3}\frac{E_p^2}{M_i \, c_s^3} \, ,
%\] 
%
%where $a$ is the lattice spacing, $\xi \sim 5-100$ fm is the coherence length of the superfluid, $M_i$ the mass of the ions,
%$E_p$ the pinning energy per nucleus and $c_s \sim 10^9$ cm s$^{-1}$ is the local speed of sound in the inner crust. 
%Using the fact that for a bcc lattice, neglecting electrons, $\rho_p = 2 M_i/a^3$ and 
%with the identification 
%\[
%E_p \, a^{-1} \, \xi^{-1} \, \rightarrow \, f_p
%\]
%we have 
%
%\begin{align}
%\label{eq:drag-phonon}
%\eta_{ph} \, \approx \, 0.16 \frac{f_p^2}{\rho_p \, \xi \, c_s^3} \, .
%\end{align} 
%
%All these quantities are not constant in the inner crust and the above formula is only an order of magnitude
%estimate for $\eta_{ph}$, that turns out to be strongly dependent to the local velocity of sound.
However, at present only rough estimates of the drag parameter in the crust can be found in the literature. 
For this reason, we decide to take as a title of example a fiducial constant value for the ``phononic'' drag parameter in the inner crust,
namely $\eta_{ph}\sim 10^2$ g s$^{-1}$cm$^{-1}$. 
This choice is made on the basis of the inferred parameters presented in \cite{HP12}, see Table 1 therein.

When the velocity of vortex lines relative to the lattice nuclei is much higher, other dissipative processes come into the play.
Following \citet{J92} in this case the dominant dissipative process involves the creation of quasiparticles (Kelvin waves) in the vortex cores. 
This phenomenon is still little studied and only rough estimates are available at the moment. However the phenomenological approach of
\citet{HP12} tells us that the drag associated to excitation of Kelvin waves is roughly seven orders of magnitude greater than the
phononic drag, thus we consider $\eta_{kw}\sim 10^9$ g s$^{-1}$cm$^{-1}$.
Figure \ref{fig:eta} shows the two full drag profiles as a function of the density given by Eq. \eqref{eq:gen-drag};
in both cases (``phononic'' and ``kelvonic'') the mechanism that dominates dissipation in the core  is given by Eq. \eqref{eq:drag-ander}.

The fact that for different vortex velocities we have to interpolate between the phononic drag and a kelvonic drag
is a direct manifestation of the approximation given in Eq. \eqref{eq:vel-y-b-w}: the power dissipated
per unit length by a vortex is not in general of the form $\eta (\bm{v}_L-\bm{v}_p)^2$ and the assumption of a velocity independent (or lag independent)
parameter $\eta$ forces the dissipated power to be quadratic in the vortex velocity. For this reason 
the prescription of a lag dependent viscous drag given in Sec. \ref{susec:effective-pinning} seems a viable choice for a dynamical simulation.
The difficulty is that to implement this ``viscous drag prescription'' in a physically reasonable way we need a better understanding of drag
and pinning.
%
%
%
%
%
%
%
%
%
%
%
%
%INPUT PROFILES%INPUT PROFILES%INPUT PROFILES%INPUT PROFILES%INPUT PROFILES%g PROFILES%INPUT PROFILES%INPUT PROFILES
%
\subsection{Numerical results}
\begin{figure}
    \centering
    \includegraphics[width=.45 \textwidth]{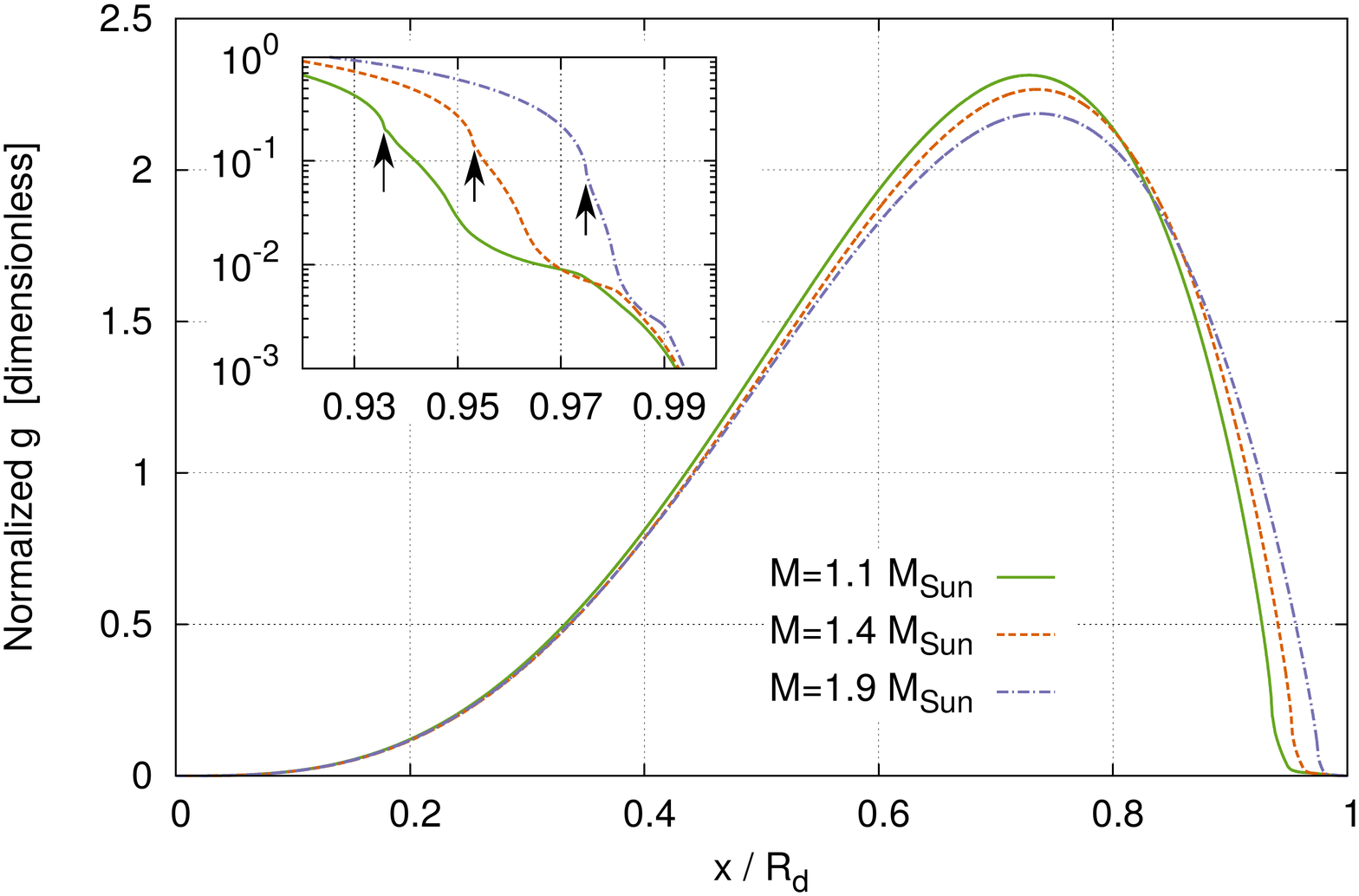}
    \caption{Dimensionless weight $g(x)$ for three neutron stars with mass 
    1.1, 1.4 and 1.9 $M_\odot$ using the SLy equation of state and its consistent superfluid neutron fraction $x_n(\rho)$.
    Here, in order to compare stars with different masses, $g$ is normalized on the interval $[0,1]$. In the insert 
    we focus on the outermost cylindrical shell and the arrows indicate the surfaces $x=R_c$ for the three masses. 
    }
    \label{fig:g}
\end{figure}
\begin{figure}
	\centering
	\includegraphics[width=.45 \textwidth]{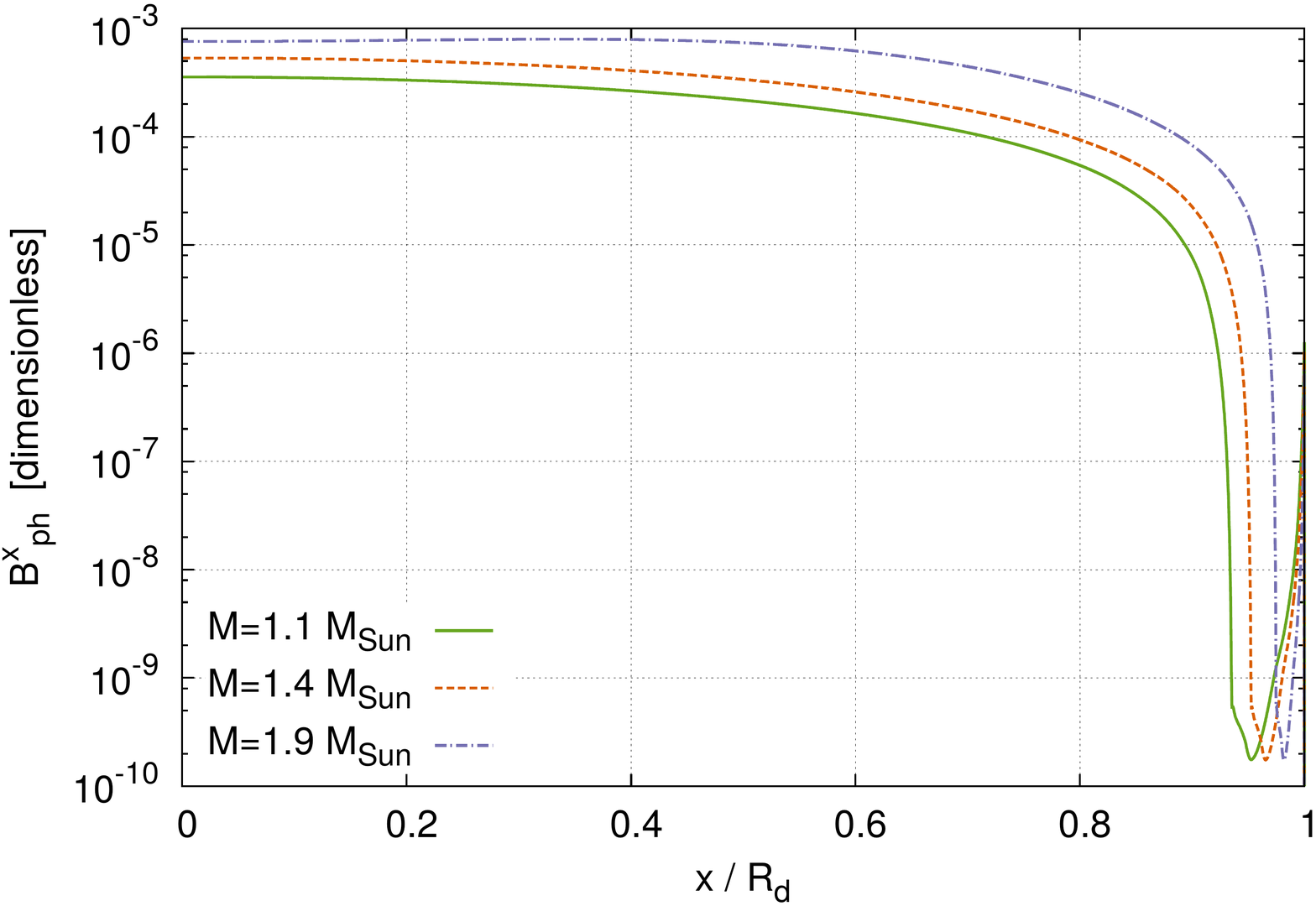}
	\caption{The dimensionless profile $\mathcal{B}^x_{ph}(x)$ for three neutron stars with mass 1.1, 1.4 and 1.9 $M_\odot$,
	using $\eta(\rho)=\eta_{ph}(\rho)+\eta_{es}(\rho)$. Note that we started from a discontinuous $\eta$ at $\rho=\rho_c$
	but the integration over the vortex length gives a continuous profile. Here we used the SLy equation of state.
	}
	\label{fig:dragph}
\end{figure}
\begin{figure}
	\centering
	\includegraphics[width=.45 \textwidth]{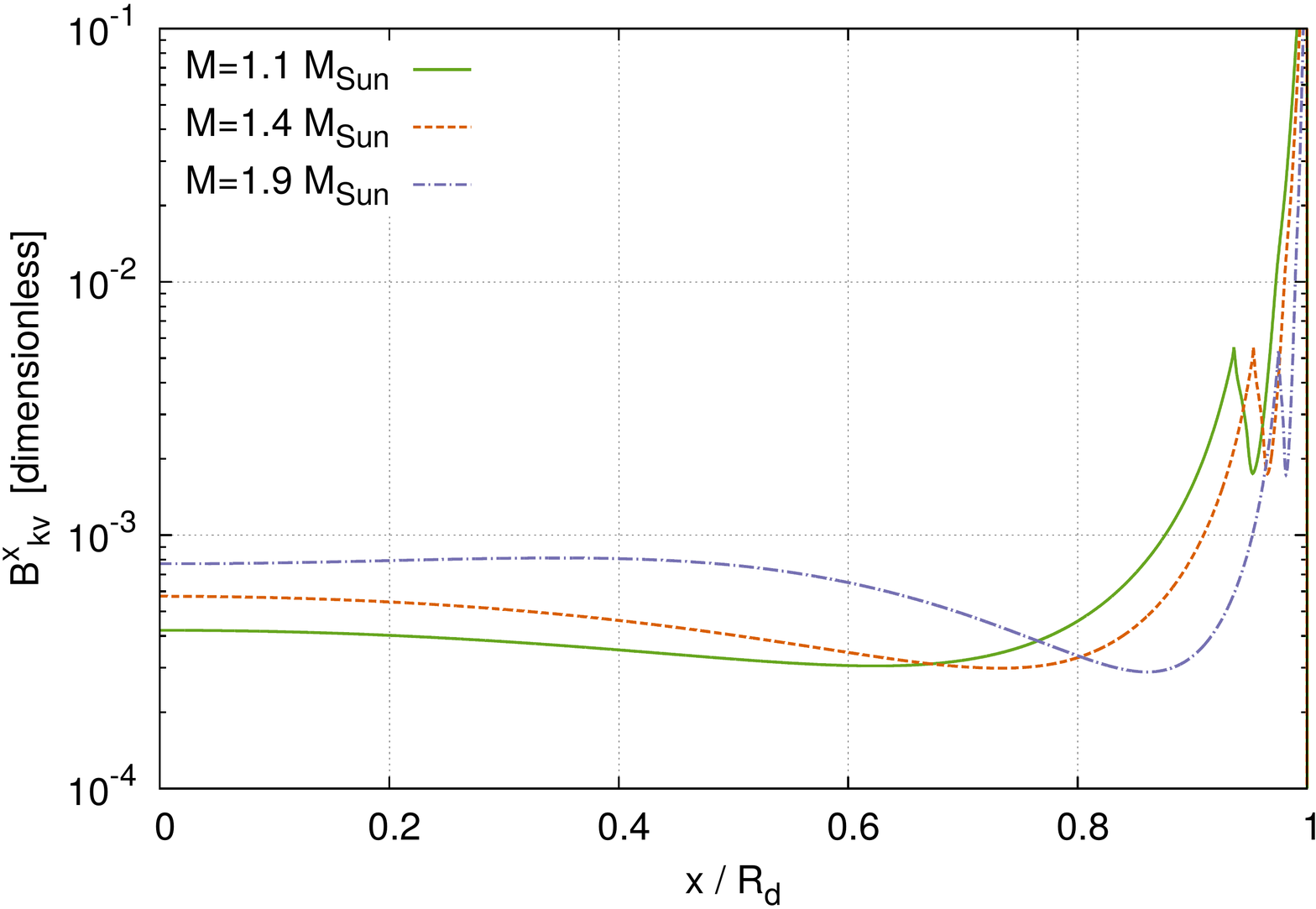}
	\caption{The dimensionless profile $\mathcal{B}^x_{kv}(x)$ for three neutron stars with mass 1.1, 1.4 and 1.9 $M_\odot$,
	using $\eta(\rho)=\eta_{kv}(\rho)+\eta_{es}(\rho)$. For vortex lines near the rotational axis the drag is almost
	entirely given by the electron scattering and there is nearly no difference with respect to the ``phonon'' case.
    Here we used the SLy equation of state.
	}
	\label{fig:dragkv}
\end{figure}
\begin{figure}
	\centering
	\includegraphics[width=.45 \textwidth]{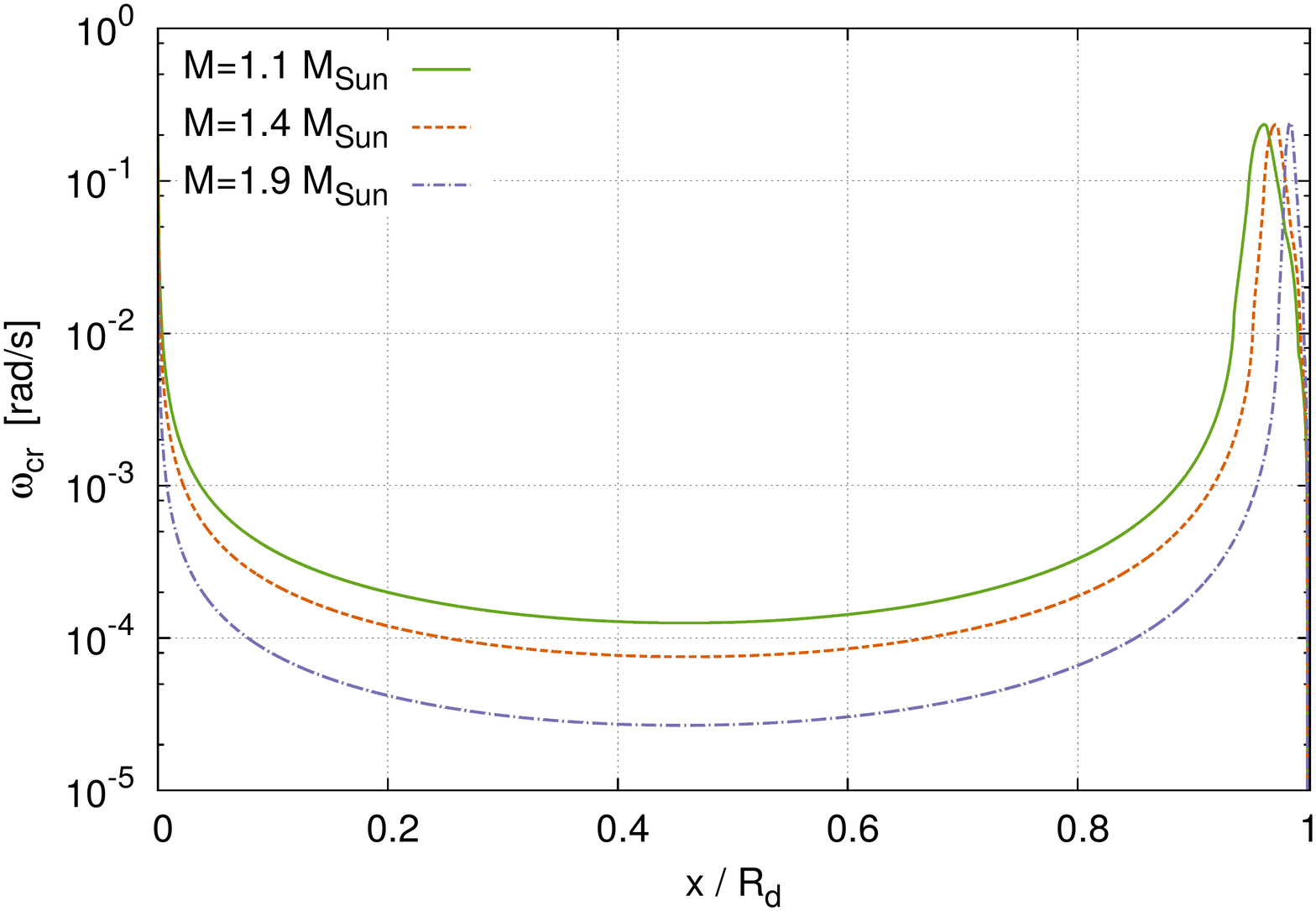}
	\caption{Critical lag profile $\omega_{cr}(x)$ for three neutron stars with mass 1.1, 1.4 and 1.9 $M_\odot$
    using the SLy EOS.
    The critical lag calculated without taking into account the entrainment effect is lower by one order
    of magnitude in the crust \citep{SP12}. 
    }
	\label{fig:wcr-sly}
\end{figure}
\begin{figure}
	\centering
	\includegraphics[width=.45 \textwidth]{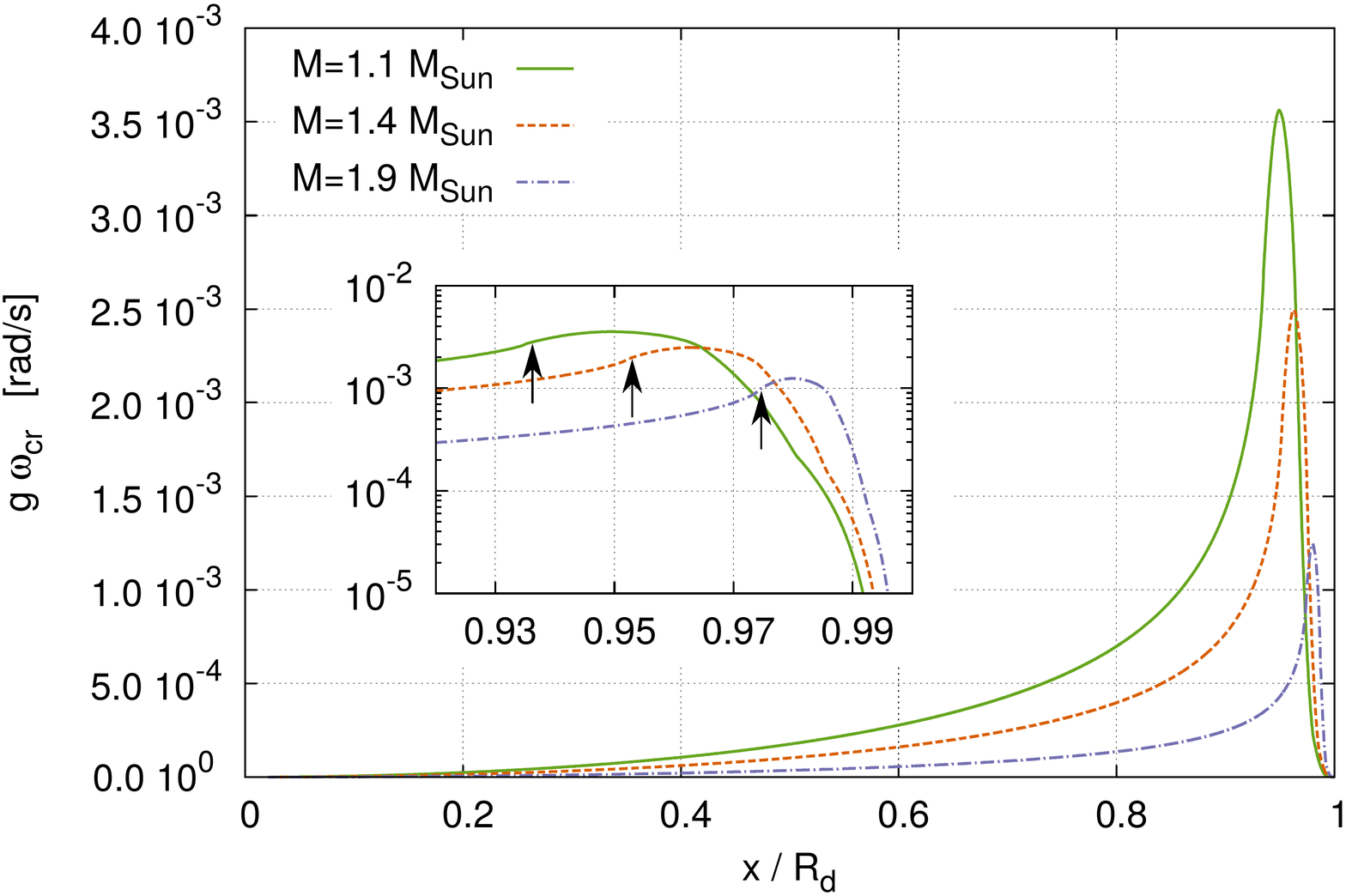}
	\caption{The product  $g(x) \omega_{cr}(x)$ is  plotted for three neutron stars with mass 1.1, 1.4 and 1.9 $M_\odot$
    using the SLy EOS. The area enclosed by each curves represents the average critical lag, $\langle\omega_{cr}\rangle$, corresponding to that mass: they are listed in Table \ref{tb:properties}. In the insert we zoom the outermost cylindrical shell and the arrows indicate the surfaces $x=R_c$ for the three masses.  
    }
	\label{fig:gwcr-sly}
\end{figure}
\begin{table}
    \centering
    \caption{Some stellar structure parameters are given for three different masses (low, medium and high).
        Neutron drip begins at the radius $R_d$ that corresponds to a density 
        $\rho_d = 1.53\times 10^{-3} \,\rho_0$. Similarly $R_{c}$ is the core radius and corresponds to $\rho_c \approx 0.47  \,\rho_0$.
        The upper panel refers to the SLy EOS. In the lower panel we use  the GM1 EOS,
        together with its consistent $x_n(\rho)$ in the core, while in the crust nothing changes with respect to the SLy case.
    }
    \setlength{\tabcolsep}{12pt}
    \begin{tabular}{@{}lrrrrr@{}}
        \hline
        $\,\,M$         & $R_d\,\,$ & $R_c\,\,$ & $I/10^{45}$  & $I_v\,$ & $q\,\,$\\
        ($\,M_\odot$) & (km)      & ($R_d$)   & (g cm$^2$)   & ($I$)   &   \\
        \hline
        1.1 &   11.41 &   0.936 &  0.86 &  0.928 &  12.9 \\
        1.4 &   11.43 &   0.953 &  1.09 &  0.941 &  16.1 \\ 
        1.9 &   10.92 &   0.974 &  1.35 &  0.958 &  22.9 \\ 
        \hline
        1.1 &   13.08 &   0.926 &  1.14 &  0.901 &  9.2  \\ 
        1.4 &   13.28 &   0.945 &  1.51 &  0.913 &  10.5 \\ 
        1.9 &   13.24 &   0.966 &  2.04 &  0.916 &  10.9 \\ 
        \hline                
    \end{tabular}
    \label{tb:strcture}
\end{table}
We use the physical input described in the previous section to construct the profiles $\mathcal{B}^x(x)$, $w_{cr}(x)$ and $g(x)$ by performing numerically
the integrations over the $z$-axis for a given cylindrical radius $x$. 
To study how the input of our model are sensitive to a change of the EOS in the core, we decided to use SLy and the stiffer GM1. 
For both cases (SLy for the whole star or GM1 in the core plus SLy in the crust), some structural parameters of the star are summarized in Table \ref{tb:strcture}
for three different masses. As expected, the stiff equation of state GM1 gives rise to smaller values of $q$.

In the following we present the results of the cylindrical profiles for the SLy case only, 
since the profiles obtained using GM1 in the core are qualitatively similar. Since we want to compare stars with different masses,
and we already remarked that the actual value of $R_d$ is irrelevant in our model, the profiles are shown rescaling the cylindrical radius to unity.
%: in all cases the weight $g(x)$ falls very rapidly to zero for $x>R_c$.
In Fig \ref{fig:g} we plot the weight $g(x)$, that is the fundamental quantity that permits to take into account 
the spherical geometry and the non-uniform structure of the star by projecting it onto a cylindrical model. 
Since our cylindrical averages are weighted by the moment of inertia, it is clear that
the behavior of $g$ near the origin is ruled by the factor $x^3$, while the drop after the maximum 
depends on the decreasing density and on the spherical shape of the star; 
then the profile of $g$ is corrected by the details of the EOS used and by the entrainment effect.
It is worth to note that the mass of the star does not have a deep impact on the overall shape of the normalized profile $g$, thanks to the fact that in
our model we are free to rescale the cylindrical radius. However by increasing the mass of the star we find that $g$ tends to weight more 
the outermost cylindrical layers for $x \gtrsim 0.9 \, R_d$. 
%We also remark that $g(x)>1$  in the cylindrical region  $0.4 \lesssim x/R_d \lesssim 0.9$, so  that 
%are more affected by these parts of the star ($x > 0.9 \, R_d$).
%Different density profiles (either due to different masses or to different EOSs)
%quantitatively change this general picture mainly by varying the weight of the outermost cylindrical shells ($x > 0.9 \, R_d$).
Moreover the cylindrical shells that are completely immersed in the crust (namely for $x>R_c$)
contribute very little to the cylindrical averages and to the total moment of inertia of the v-component, as can be seen in the insert of Fig \ref{fig:g}. 

Figures \ref{fig:dragph} and \ref{fig:dragkv} show the $\mathcal{B}^x$ profile calculated with electron scattering in the core and 
either phononic or kelvonic drag in the crust. 
We label the total dimensionless drag $\mathcal{B}^x$ in these two cases as  $\mathcal{B}^x_{ph}$  and  $\mathcal{B}^x_{kv}$.
The dominant contribution from electron scattering in the core makes these two profiles quite similar
for $x \lesssim 0.6 \, R_d$. 
Notice that switching from phononic drag to kelvonic drag in the crust is relevant
also for vortex lines that are not completely immersed in the crust ($0.8 \, R_d \lesssim x < R_c$).

The critical lag $\omega_{cr}$ given by Eq. \eqref{eq:cr-lag} is shown in Fig. \ref{fig:wcr-sly} for the SLy EOS. 
When GM1 is used in the core the result is qualitatively similar: the critical lag profiles, even in presence of entrainment 
still have the typical peak of the ``snowplow-model''. This peak gets closer to the cylindrical shell $x=R_d$ as we increase the mass of the star. 
When GM1 is used in the core, the main difference with respect to the SLy case is the higher value of the central plateau between $0.2\lesssim x/R_d \lesssim0.8$. 

Finally, in Fig. \ref{fig:gwcr-sly} we show the product $g(x) \omega_{cr}(x)$ for the SLy EOS; by definition, the area enclosed by each curve is the mean critical lag $\langle\omega_{cr}\rangle$, listed in Table \ref{tb:properties}. It is worth noting that the main contribution to this comes from regions with $x > 0.6\, R_d$, which do not contain the possibly exotic inner core: the cylindrical angular momentum reservoir is spread across  the inner crust and the outer core of the neutron star.
\section{An application: estimates of the angular momentum reservoir and of the spin-up timescales}
\label{subsc:max-mass}
Here we briefly show how some of the results developed in the previous section can be useful to make quantitative estimates
even without solving explicitly the equations of motion \eqref{eq:final-sys-v} and \eqref{eq:final-sys-p}.
%\par 
%In the previous section we obtained the critical lag profiles for three different masses both for SLy and GM1 equations of state.
% In doing this we choose to use one of the results of \cite{SP15} for the input $f_p(\rho)$ and this has strong repercussion 
%on the related critical lag profile. But again we stress here that this is only a way to show a line of reasoning.

Firstly we give a rough estimate of the spin-up timescale:
we assume that the outward velocity after triggering a large glitch event is given by Eq. \eqref{eq:vel-y-b-w}
and that all vortex lines unpin so that $Y\sim1$ everywhere. 
Thus the rigid approximation of Sec. \ref{subsc:reduction-rigid-model}
implies that the spin-up timescale can be defined as
\begin{equation}
\label{eq:spin-up-time}
\tau_{kv} = \frac{P}{4 \pi \langle \mathcal{B}^x_{kv} \rangle (1+q)}\, ,
\end{equation}
in complete analogy with Eq. \eqref{eq:relax-time}: the body averaged relaxation time is the weighted harmonic mean of the local 
relaxation times. This is reliable as long as repinning, or a significant change
in the value of the friction functional, occurs for times greater than $\tau_{kv}$ since the beginning of the avalanche.
Similarly we can calculate $\tau_{ph}$, that is the same quantity but using $\mathcal{B}^x_{ph}$.
We list these values (in units of the pulsar period $P$) in Table \ref{tb:properties}.
These values are only global quantities, intimately related to the rigid approximation described in Sec 3.1. 
Moreover the listed spin-up timescales are referred to the extreme situation where all the vortex lines
are unpinned ($Y=1$). For the Vela pulsar, using a period $P \sim 0.09\,$s, all the listed spin-up timescales are less than a minute and are
particularly fast for a soft EOS like SLy. They are definitely smaller than the present observational ``blind window'' of about 40 s \citep{DM02}, thus pointing to fast dynamical transients that need a much better timing resolution to be revealed: the first few seconds of a glitch event may provide a direct insight into the dynamics of superfluid vortices inside a pulsar.

\begin{figure}
    \centering
    \includegraphics[width=.4 \textwidth]{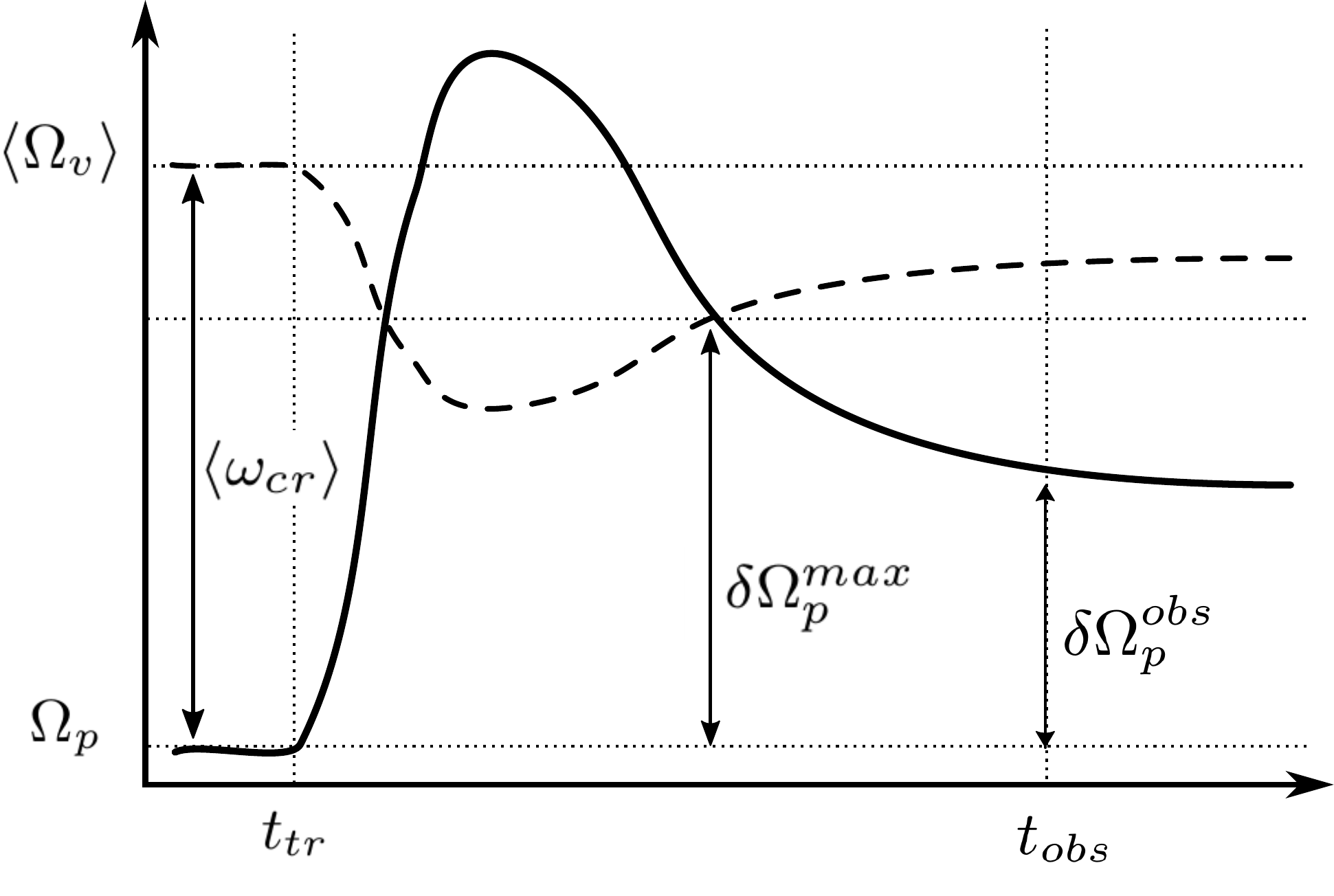}
    \caption{Sketch (out of scale) of the expected behavior of a typical glitch, with the unresolved region where an overshoot can occur and the maximal glitch $\delta\Omega^{max}_p$ defined in the text.   The time at which the glitch is triggered is labeled with $t_{tr}$, whereas $t_{obs}$ is the earliest time that can be
        resolved by current observations. 
    }
    \label{fig:glitch-sketch}
\end{figure}

To give an upper limit to the angular momentum reservoir, we must estimate the theoretical  maximal glitch for a given EoS and for different masses. In turn, this can be compared to observations in order to constrain the pulsar mass.
%, by imposing that $\Omega_v = \Omega_p+\omega_{cr}$
%and that all the angular momentum reservoir of the lag is released to the $p$-component. 
Conservation of  total angular momentum before and after the glitch reads 
\[
I \Omega_p+I_v\langle \omega_{pre} \rangle\,=\,I (\Omega_p+\delta\Omega_p)+I_v\langle \omega_{post} \rangle 
\]
in terms of the initial and final average lags. We can define a {\em maximal} glitch amplitude by taking $ \langle \omega_{pre} \rangle=\langle\omega_{cr}\rangle$ and $\langle \omega_{post} \rangle =0$, namely a glitch where initially the whole star is just subcritical (in the snowplow scenario, this maximizes the angular momentum reservoir) while after the glitch the average lag is reduced to zero and the reservoir is ``empty". We thus obtain
\begin{equation}
\label{eq:max-glitch}
\delta \Omega^{max}_p \, = \, \frac{I_v}{I}\,\langle \omega_{cr}(x) \rangle  \, .
\end{equation}
Note that we don't require that after the glitch the lag is everywhere null, but only that the lag is null in the mean at a given instant 
after the triggering of the avalanche has occurred. Although this maximal glitch does not correspond to a real situation and is probably never realized dynamically, it still sets an upper limit to the observed glitch amplitude. Indeed, 
 a sketch (out of scale) of a typical large glitch event is shown in Fig \ref{fig:glitch-sketch}; it is clear that 
$\delta\Omega^{max}_p$ is not really the maximum value of the frequency jump, but rather it is the difference between the value of $\Omega_p$ when the mean lag is 
instantaneously zero and the initial $\Omega_p(t_{tr})$. An overshoot of the p-component is expected since $q\gg 1$. Now, the very short spin-up timescales that we find by putting $Y=1$ suggest that for a large glitch (where all vortices have been unpinned thus maximizing the avalanche) the overshoot is realized on timescales comparable to $\tau_{kv}$, namely a few seconds for Vela.
Thus if we observe a glitch amplitude $\delta\Omega_{obs}$ about a minute after the triggering event (say at time $t_{obs}$, 
corresponding to the observational blind window) we can be confident that
we are observing the glitch amplitude after the condition of ``null lag'' has occurred (see Fig \ref{fig:glitch-sketch}), so that 
$\delta\Omega^{max}_p$ is actually un upper limit for the observed glitch.  
Conversely, any dynamical simulation will give a glitch amplitude (measured at times well after the short-time transients) 
which is smaller than $\delta\Omega^{max}_p$; therefore only stellar structures that permit to obtain a $\delta\Omega^{max}_p$ greater than the largest 
observed glitch are physically acceptable for a given pulsar.
Some values of $\delta \Omega^{max}_p$ are listed in Table \ref{tb:properties} for different EOS and stellar masses.
For example, looking at the listed values of $\delta\Omega_p^{max}$, 
we can conclude that a mass of $1.4 \, M_\odot$ is enough to reproduce the maximum observed glitch of the Vela ($\approx 2.1\times10^{-4}\,$rad/s)
when the SLy EoS is used. With more values than those listed we find the more stringent limit $M_{Vela}<1.45 \, M_{\odot}$.
The stiffer GM1 allows for  larger masses but still rules out the possibility that the Vela is a very massive neutron star ($M_{Vela}<1.8 \, M_{\odot}$).
These upper bounds to the mass of the Vela can only be lowered by the observation of future, even larger glitch events. 
Note that the identification of the Vela as an intermediate-mass neutron star has also been proposed on the basis of its 
peculiar surface cooling \citep{YP04}.

Finally we note that the maximum glitch amplitude $\delta \Omega^{max}_p$ does {\em not} depend on entrainment, as can be seen 
directly from its definition in Eq. \eqref{eq:max-glitch}. 
Our conclusion is thus robust against the uncertainties of the entrainment parameters.

%We can give a generalization of \eqref{eq:max-glitch}: in general,
%given whatever lag profile $\omega(x)$, the associated maximum glitch amplitude is 
%$\delta \Omega^{max}_p[\omega] \, = \, \langle \omega \rangle_x \, I_v/I$ and the interpretation is exactly the same of the previous case.
%

\begin{table}
    \centering
    \caption{Timescales and maximal glitches as a function of stellar mass are calculated using the SLy EOS (upper panel)
        and the GM1 EOS with its consistent $x_n(\rho)$ in the core (lower panel).
        The phonon and kelvon relaxation time are obtained using  Eq. \eqref{eq:spin-up-time} and are given in units of the pulsar period $ P $. }
    \setlength{\tabcolsep}{12pt}
    \begin{tabular}{@{}lrrrr@{}}
        \hline
        $\,\,M $         
        & $\tau_{ph}\,\,$
        & $\tau_{kv}\,\,$
        & $\langle \omega_{cr}\rangle\,\,$ 
        & $\delta \Omega^{max}_p\,\,$
        \\
        ($\,M_\odot$)
        & ($P$)      
        & ($P$)   
        & ($10^{-4}$rad/s)
        & ($10^{-4}$rad/s)     
        \\
        \hline
        1.1 &  62.1  &  13.2  &  4.17 & 3.87  \\
        1.4 &  29.2  &  12.1  &  2.51 & 2.37  \\ 
        1.9 &  9.0   &  7.2   &  0.92 & 0.89  \\ 
        \hline
        1.1 &  445  &  13.9  &  6.99  & 6.30  \\ 
        1.4 &  261  &  17.1  &  4.37  & 3.92  \\ 
        1.9 &  136  &  26.3  &  2.04  & 1.87  \\ 
        \hline
    \end{tabular}
    \label{tb:properties}
\end{table}
Of course we can repeat this line of reasoning for all pulsars that display large glitches, but
a detailed discussion and improvement of the estimate of the angular momentum reservoir will be given in a dedicated work. 
Here our aim is just to propose that the largest glitch observed in each pulsar can 
put a constraint on the pulsar mass. This approach to constrain some star properties through glitch observations is very different 
when compared to  the estimate of the moment of inertia fraction using the activity parameter  \citep{LE99,AG12, C13}; 
indeed, the activity parameter is a mean quantity over many decades of pulsar evolution, while here we take the largest observed event, 
which may be a better indicator of the pulsar's glitching strength.
\section{Conclusions}
\label{sec:conclusions}
In this paper we have developed a  dynamical model for superfluid pulsars, that 
can take into account consistently  the layered structure of the star, the differential rotation of the superfluid and the presence of strong density-dependent entrainment. 
In our treatment we tried to point out that all the complex behavior of the rearrangement of  the vortex configuration 
is hidden into the functional $\mathcal{B}$, introduced in Eq. \eqref{eq:general-B}.
In our estimate of the Vela maximal mass and of the spin-up timescales we used the simple form of $\mathcal{B}$ given 
by Eq. \eqref{eq:vel-y-b-w}. However it is always possible to refine the model with better proposals for $\mathcal{B}$ and to 
study the consequences on the dynamics of the observable variable $\Omega_p$.

The main result of the paper is  the set of dynamical equations, Eqs.\eqref{eq:final-sys-p} and \eqref{eq:final-sys-v}, that 
constitute the mathematical framework for future simulations and modeling of pulsar glitches. 
They describe the exchange of angular momentum between two effective components, characterized by moments of inertia $I_v$ and $I_p$, determined by entrainment,  and whose angular velocities  are  $\Omega_v(x,t)$ and the observable $\Omega_p(t)$.
Even without solving explicitly the equations, we were able to draw some general quantitative conclusions; in particular, 
we have  sketched a new method to estimate an upper bound to the mass of a large glitcher using observations.
Moreover our conclusion on the mass of the Vela pulsar is robust against the uncertainties of the entrainment estimates,
as can be seen from Eq. \eqref{eq:max-glitch}.

A central feature of the quantitative example described in the previous section concerns the nature of superfluid vortices extending across the crust-core interface. 
As explained before, we consider that the change in the symmetry of the order parameter (from the S-superfluid to the P-superfluid)
may affect the vortex core structure but doesn't decouple the core superfluid and the crustal superfluid.
In most of the existing literature the reservoir of angular momentum is instead restricted to the crustal superfluid. 
However \cite{C13} and \cite{AG12} pointed out independently that with the inclusion of entrainment the superfluid into the crust has not enough mobility 
to fit the activity parameter of many pulsars (in particular the large glitchers, like the Vela). In our model with continuous vortices, 
also the vortex lines that are not totally immersed in the crust ($x<R_c$) play an important role and the reservoir of angular momentum 
is enough to reproduce the largest glitch observed in the Vela, provided its mass is not too large. 
A more detailed study of the parameter space for the different input quantities
as well as simulations of glitches induced by a massive unpinning of vortex lines will be the issue of future work. 
Of course, this kind of simulation makes more sense if it can eventually be compared to better resolution timing data: 
the first few seconds of a glitch with their massive exchanges of angular momentum may be far more revealing and constraining 
for the star structure than the later post-glitch evolution.
%Here we just reported, in Tables \ref{tb:strcture} and \ref{tb:properties}, some non-dynamical (averaged) properties of a star for three different masses and a given equation of state (SLy or GM1).
%In particular the listed relaxation times $\tau_{kv}$ and $\tau_{ph}$, found by means of the rigid-approximation of Sec. \ref{subsc:reduction-rigid-model}, are consistent with the fast unresolved spin up ($< 60\,$s). Thus a future resolution of a glitch event should give us precious informations about the functional form of $\mathcal{B}$. Hopefully this also will put another constraint on a pulsar mass, via the comparison with the measured spin up of dynamical simulations. 
%We conclude with an open question: is it possible to find a particular and reasonable form for $\mathcal{B}$
%such that a simulation of the dynamical system proposed will display a SOC behavior?
%At the moment, given the values in Tab.\ref{tb:properties}, we can only notice that high mass pulsars with a soft EOS should have a steeper spin up. 
%A stiff EoS on the other hand (like GM1) would produce a mild spin up.

Finally we remark that despite the early conclusion of \citet{RS74}, the analogy with laboratory experiments on superfluid Helium tell us that turbulence is likely to develop in some internal regions of neutron stars. 
However there is far from a consensus as to which regimes will lead to turbulence and how this will develop, and all the
work on this topic is at the moment very exploratory. 
Thus the scope of our paper is to clarify how superfluid entrainment affects the macroscopic dynamics in the simplest situation of straight vortices and to test quantitatively the consequences of such a widespread assumption.
In fact, all vortex-creep models (since the seminal work of \citet{AA84}) are derived under the simplification of straight vortices and do not include the effect of entrainment or stratification.  Similarly, models based on the two-fluid formalism with mutual forces  have either neglected entrainment \citep{HP12} or have been studied under the assumption of uniform entrainment and rigid rotation of the neutron superfluid \citep{PC02,SP10}.
%In our model, instead,  differential rotation, density-dependent entrainment, stellar stratification and vortex dynamics are consistently accounted for;  
%moreover, it provides a dynamical realization of the static ``snowplow model''
%\citep{P11,SP12} and clarifies how to introduce entrainment in the evaluation of the critical lag profile $\omega_{cr}(x)$. 
%
%We conclude by stressing that, despite 
%the caveats on our working assumptions, it is important to look for simplified but clear prescriptions to create body-averaged models.
%Here we tried to make a first step in this direction by assuming straight vortex lines in a Newtonian framework 
%(for simplicity in our numerical examples we used the TOV equations for the structure of the star, but the classical moments of inertia).
At this point it is quite simple to introduce general relativistic corrections for slow rotation, and to test different EoS and alternative input profiles for pinning, 
entrainment and drag.
\section*{Acknowledgments}
Partial support comes from NewCompStar, COST Action MP1304.  
%We thank  Stefano Seveso and Bryn Haskell for helpful discussions. 
%MA also thanks Bryn Haskell for his hospitality at the 
%University of Melbourne during the writing of the manuscript.
%
\nocite{*}
\bibliographystyle{mn2e}

\bsp

\label{lastpage}

\end{document}